# Modeling the impact of control zone restrictions on pig placement in simulated African swine fever in the United States

Chunlin Yi[1], Jason A. Galvis[1], Gustavo Machado[1,2*]

[1] Department of Population Health and Pathobiology, College of Veterinary Medicine, North Carolina State University, Raleigh, North Carolina, USA.

[2] Center for Geospatial Analytics, North Carolina State University, Raleigh, NC, USA.

***Corresponding author:** gmachad@ncsu.edu.

## Abstract

African swine fever (ASF) is a highly contagious viral disease that poses a significant threat to the swine industry, requiring stringent control measures, including movement restrictions that delay pig placements, impacting business continuity. The number and economic impact of unplaced healthy animals due to control zone restrictions remains unmeasured. This study evaluates the economic and epidemiological impacts of control zone placement restrictions during simulated ASF outbreaks in U.S. commercial swine farms. We model the spread of ASF and apply the U.S. National Response Plan (NRP) alongside alternative mitigation strategies, analyzing key metrics such as the number of unplaced pigs, depopulated pigs, infected farms, and total economic losses. Our findings estimate the median number of unplaced pigs in the first year was 153,020 (IQR: 27,377–1,307,899) under the NRP scenario. Shorter control zone durations (20–25 days) effectively reduce the median number of unplaced pigs by 16.7% to 33.5%, whereas longer durations (40 days) increase unplacement numbers by 32%. The median number of depopulated pigs remains broadly consistent across all durations. Expanding the infected zone (5–15 km) increases the median number of unplaced pigs by 53.6%–282% while reducing depopulated pigs by 28.8% to 73.9%, respectively. Economic losses are estimated through a model that includes depopulated and unplaced animals requiring culling. We examined the situations when 5%, 12%, or 20% of unplaced pigs required culling and found that the total cost ranged from zero (no second infection) to over $800 millions.

## 1. Introduction

African swine fever (ASF) is a contagious viral disease affecting domestic and wild pigs (Dixon et al., 2019; Galindo and Alonso, 2017). While ASF poses no direct threat to human health, its consequences on pig populations and the economy are catastrophic (Herrera-Ibatá et al., 2017; Mason-D'Croz et al., 2020; Niemi, 2020). From its identification in East Africa in the early 1900s, the virus has since spread to multiple countries across Africa, Europe, and Asia. It has been detected in the Caribbean islands of Haiti and the Dominican Republic (Chenais et al., 2019; Gonzales et al., 2021; Jean-Pierre et al., 2022; Mighell and Ward, 2021; Schambow et al., 2025). The proximity of ASF to the U. S. has raised significant concerns about the potential for the disease to cross into North America (Herrera-Ibatá et al., 2017; Jurado et al., 2019; Schambow et al., 2022).

The between-farm transmission of ASF occurs through direct contact, indirect contact, and local spread (Deka et al., 2024; Guinat et al., 2016; Hayes et al., 2021; Mazur-Panasiuk et al., 2019; Olesen et al., 2020, 2017). Direct transmission often happens when healthy pigs come into contact with infected animals' blood, tissues, or bodily fluids (Olesen et al., 2017). Indirect transmission is facilitated through contaminated surfaces, such as fomites, trucks, feed, or clothing worn by farm workers (Mazur-Panasiuk et al., 2019). In addition, mechanical vectors like soft ticks (Ornithodoros species) and biting flies may spread the virus between animals and farms (Bonnet et al., 2020; Golnar et al., 2019). Wildlife has also played a role in farm-to-farm transmission, with the most risk posed by wild pigs (Pavone et al., 2023). The movement of people, animals, and equipment between neighboring farms further exacerbates local spread, making biosecurity measures essential to preventing introduction and further spread (Vergne et al., 2017). Numerous studies have analyzed the effectiveness of control strategies for ASF (Guinat et al., 2017; Sykes et al., 2023; Reichold et al., 2022; Machado et al., 2022; Lamberga et al., 2024). Guinat et al. demonstrated that culling ASF infected herds and movement bans for neighboring herds were the most effective intervention strategies among tested scenarios (Guinat et al., 2017). Similarly, Sykes at al. (Sykes et al., 2023) demonstrated that implementing a control strategy combining quarantine, depopulation, movement restrictions, contact tracing, and enhanced surveillance was estimated to reduce secondary cases by 79%. While these strategies are expected to contain the virus and facilitate early detection, they also present significant economic challenges. Animal movement standstill and the high costs of surveillance and biosecurity measures are among the most pressing concerns (Halasa et al., 2018, 2016). Halasa et al. examined the impact of control zone size and duration on the outcomes of a hypothetical ASF epidemic in Denmark (Halasa et al., 2018). Their findings revealed that prolonged control zone durations significantly affect trade and economic outcomes without yielding substantial epidemiological benefits. Increasing control zone sizes provided a limited advantage for small to moderate outbreaks, proving more useful only in large-scale epidemics (Halasa et al., 2018).

One of the most critical economic and logistical challenges arises from movement restrictions from and to control zones (Otte et al., 2004). Most ASF control and eradication policies restrict pig movement from and to control zones. Healthy pigs within control zones cannot be transported to other farms or, in some cases, even to slaughterhouses without testing and movement permission (USDA, 2023). In addition, pigs outside control zones are also not allowed to enter these areas. This disruption hinders the production chain and creates significant welfare and economic issues (East et al., 2014; Tildesley et al., 2019). Over time, restricted movements cause farm overcrowding, increasing stress, tail biting, and disease spread among pigs (Cohen, 2010; Kritas and Morrison, 2007). Ultimately, the lack of available space forces culling healthy pigs (VanderWaal et al., 2021). Therefore, restrictions on pig movements within control zones during ASF outbreaks represent a significant challenge for the industry.

This study aimed to simulate ASF outbreaks where control strategies were implemented based on the USDA's National Response Plan (NRP) (USDA, 2023) to estimate impacts of varying control zone sizes and durations on the number of unplaced pigs, the number of infected farms, and the number of depopulated pigs. The model incorporating depopulation and unplacement costs was proposed to assess the total economic losses. The findings of this work could guide more adaptive policy decisions, optimizing control zone size and duration to minimize economic losses while maintaining disease control.

## 2. Materials and methods

### 2.1. Population, animals, and trucks movement data

The dataset used in this study comprises commercial swine farm population and geolocation data of 1,981 farms, as well as swine and vehicle movement records from three swine production companies with operations in three U.S. states. The population data includes geolocation information, premises identification, production types (sow, nursery, finisher, gilt development, wean-to-finish, and farrow-to-finish), and farm capacity (Cardenas et al., 2024). The movement data of pigs and trucks spans from January 1st, 2020, to December 31st, 2020, capturing the purpose of each movement, dates, origin and destination farms, the number of swine transported, and the type of vehicle used. For more details on the movement data and description of the pig movement network, see Galvis et al. (Galvis and Machado, 2024) and Cardenas et al. (Cardenas et al., 2024).

### 2.2 Mathematical model

We utilized our previous SEID model (Sykes et al., 2023; Yang et al., 2020; Zhang et al., 2021) (Figure 1) to generate ASF epidemics. Briefly, the probability of transition from S to E compartment, denoted as $Y_{it}$, is determined by the force of infection from movements of exposed pigs, movements of contaminated vehicles, and local transmission (Sykes et al., 2025). The movement of pigs from an infectious or detected farm contributes to the probability of farms transitioning into the I or D compartments, represented by $V_{it}$ and $W_{it}$, respectively. Farms transition from compartment E to I through the development of infectiousness rate of $1/\sigma$ where $\sigma$ is the latent period of the ASF virus. The transition from the I to the D compartment is based on the detection of infection, modeled by a logistic function $f(x) = \frac{L}{1 + e^{-k(x-x_0)}}$, where L represents the effectiveness of surveillance for the production type, k is the logistic growth rate, x is the time spent in the I compartment and $x_0$ is the midpoint time for detection (Sykes et al., 2023). We assumed $x_0 = 10$ days (Malladi et al., 2022) before the first detected case, and $x_0 = 7$ days after the first farm was detected, reflecting increased awareness.

We constructed a daily, farm-level, three-layer contact network to simulate the dynamics of ASF transmission among farms (Sykes et al., 2023; Yi et al., 2022). The network models direct contact through swine movements, indirect contact via vehicle movements, and local contact between nearby farms using a gravity model that accounts for farms' spatial distribution and proximity to estimate the likelihood of local transmission events (Sykes et al., 2025).

Direct-contact networks are created using the daily between-farm swine movement data, with edges unweighted and directed according to the number of movements (Sykes et al., 2025). The forces of infection by swine movements were modeled based on the infection status of the originating farm, as follows $\lambda_{eit} = \beta_e * N_{eit}$, $\lambda_{iit} = \beta_i * N_{iit}$, $\lambda_{dit} = \beta_d * N_{dit}$, where $\lambda_{eit}$ represent routes from exposed farms to susceptible farms, $\lambda_{iit}$ routes from infected farms to susceptible or exposed farms, and $\lambda_{dit}$ routes from detected farms to susceptible, exposed, or infected farms. Here, $\beta$ denotes the transmission rate and $N$ indicates the number of movements for each route.

Indirect-contact networks are constructed based on the movement of pig trucks, market trucks, feed trucks, and other vehicles (Galvis and Machado, 2024). Their force of infection is denoted as $\lambda_{pit}$, $\lambda_{mit}$, $\lambda_{fit}$, and $\lambda_{oit}$. Vehicle visits are determined using GPS locations and the outlines of farm perimeter buffer areas (PBAs) from Secure Pork Supply plans (Pudenz et al., 2019). A vehicle is considered to have visited a farm if it remains within 50 meters of the PBA while stationary for at least five minutes. Vehicles are considered contaminated after visiting a simulated ASF-positive farm, and the force of infection for vehicle movements is weighted by pathogen stability $M_{ijt}$, which is a function of travel time and average environmental temperature (Galvis and Machado, 2024), $\lambda_{xit} = \beta_x * M_{ijt} * N_{xit}$ where $x$ presents the type of truck movement. A vehicle is considered to have visited a cleaning station if it remains stationary within a 500-meter radius of a cleaning station for at least 60 minutes, with the cleaning effectiveness assumed to be 50% (Sykes et al., 2025).

Local contact networks are constructed using a gravity model to define a transmission kernel, where the probability of infection decreases with distance. The infection probability of the local transmission $P_{lit}$ is given by the equation $P_{lit} = 1 - \prod_{i=1}^{n} \ (1 - \phi e^{-\alpha dij})$, where $i$ represents an infectious farm, $j$ a susceptible farm, and $d_{ij}$ is the distance between farms. Parameters $\phi$ and $\alpha$ control the maximum transmission probability and the steepness of the probability decline with distance (Sykes et al., 2025).

Therefore, the transition probabilities are given as $Y_{it} = 1 - e^{-(\lambda_{eit}+\lambda_{pit}+\lambda_{mit}+\lambda_{fit}+\lambda_{oit})} + P_{lit}$ , $V_{it} = 1 - e^{-\lambda_{iit}}$, and $W_{it} = 1 - e^{-\lambda_{dit}}$. The parameters used in this study were calibrated in a previously published study of our group (Sykes et al., 2025). Their descriptions and values are listed in Supplementary Material Table S1.

To align with current mitigation strategies, we incorporate five control and eradication actions outlined in the ASF National Response Plan (NRP) (USDA, 2023): (i) quarantine and depopulation of detected farms, (ii) a 72-hour movement standstill, (iii) contact tracing of direct and indirect contacts, (iv) depopulation of direct contacts, and (v) implementation of control zones and surveillance zones (Sykes et al., 2025). Detected farms are quarantined and scheduled for depopulation, assuming a daily depopulation limit of four farms and including delays based on the number of farms awaiting depopulation (Sykes et al., 2025). Post-depopulation, farms remain empty for 30 days for cleaning and disinfection before repopulation. A 72-hour movement standstill is enacted upon the initial detection of an ASF-positive farm, restricting live swine movements, including those by pig trucks, market trucks, and undefined trucks. Farms identified through contact tracing of swine and vehicle movements are quarantined for 15 days, during which they undergo regular ASF testing (Galvis et al., 2024). Only farms testing positive are depopulated, while direct contacts are quarantined but not depopulated. Control zones include a three-kilometer infected zone (IZ) surrounding infected farms and a two-km buffer zone (BZ) surrounding IZ, where farms are quarantined until they receive a negative ASF test result. Farms within these zones are subject to continuous testing, and the control and surveillance measures are lifted after 30 days if there are no new cases (Sykes et al., 2025). In the IZ, farms are tested every three days for the first two tests and every six days after that, while farms in the BZ are tested every six days (Sykes et al., 2023). Farms within control zones must test negative for ASF three and one day before moving pigs, referred to as pre-permit testing. Pigs are prohibited from moving if farms fail to meet this requirement or test positive for ASF (Sykes et al., 2025). Pig movements from farms outside control zones are restricted to enter control zones. The movement restrictions are the same for both the IZ and BZ (USDA, 2023). In this study, the daily processing capacity for blood samples was limited to 25 farms in North Carolina, ten in Virginia, and ten in South Carolina, based on laboratory capacities (Galvis et al., 2024; State Animal Health Official, 2023).

## 2.3 Alternative control scenarios

In addition to the NRP scenario, we simulate six alternative scenarios, modifying control zone duration and extension. In scenarios 1 to 3, control measures follow the NRP but with modified control zone durations—reduced to 20 days (scenario 1) and 25 days (scenario 2), while scenario 3 extends the duration to 40 days. Given that control zones may be determined on a case-by-case basis (USDA, 2023), factoring in epidemiological conditions, wild pig population distribution, and natural boundaries that could influence disease spread, scenarios 4 to 6 focus on expanding the infected zone (IZ) radius to 5 km, 10 km, and 15 km, respectively, while maintaining a 30-day control zone duration. Table 1 summarizes the parameters for each scenario, including the duration of control zones (in days) and the radius of the infected zone (in kilometers) for the NRP and six alternative mitigation strategies.

## 2.4 Economic losses from depopulation and unplacement

To quantify the economic impact of depopulation and placement restriction of pigs during ASF outbreaks, we propose a cost model that calculates the total economic loss as follows:

$$L = D + x \cdot U, \qquad 0 < x < 1$$

where $L, D,$ and $U$ represent total economic loss, the total price of depopulated pigs, and the total price of unplaced pigs, respectively. The parameter $x$ represents the percentage of unplaced pigs that ultimately require culling, reflecting the proportion of unplaced pigs that cannot be effectively utilized, sold, or redirected within the supply chain. The total depopulation ($D$) cost is calculated by summing the prices of depopulated pigs across farm types ($i$): $D = \sum_i n_i \cdot p_i^d$, where $n_i$ is the number of depopulated pigs from farm type $i$, and $p_i^d$ is the market price for a depopulated pig from farm type $i$ (Table 2). Similarly, the total price for unplaced pigs $U = \sum_i m_i \cdot p_i^u$, where $m_i$ is the number of unplaced pigs from farm type $i$, and $p_i^u$ is the market price for an unplaced pig from farm type $i$.

To assess economic losses ($L$), we incorporated variability in depopulation ($p_i^d$) and unplaced ($p_i^u$) cost across different farm types to better approximate real-world conditions (Table 2). The economic losses associated with depopulation and unplacement vary significantly even for the same farm type. For example, depopulating a sow farm typically involves culling all mature breeding sows among the most economically valuable animals in the herd. In contrast, unplacement from a sow farm primarily affects piglets of significantly lower total value; the economic loss associated with unplacement from sow farms is relatively less severe.

2.5 Outputs

For each scenario, we conducted simulations where the ASF was seeded on a single farm at the start of each run, and the spread of the virus among farms was simulated over 366 days. Each farm was seeded 100 times (Sykes et al., 2025). The key outcomes of the model include i) the number of unplaced pigs into control zones; ii) the number of infected farms; iii) the number of depopulated pigs; and iv) the epidemic elimination rate, which highlights the proportion of simulations where the epidemic is eliminated by the end of simulations. We calculated the number of unplaced pigs by control zone type (infection zone and buffer zone), production company, and farm type. Similarly, we categorized the number of depopulated pigs based on company and farm type to analyze depopulation patterns and further understand the severity of the outbreaks.

## 3. Results

### 3.1. The National Response Plan scenario

The results estimate a median of 153,020 unplaced pigs, with substantial variability indicated by a wide interquartile range of 27,377 to 1,307,899 and a maximum value of 4,556,749 (Supplementary Material Table S2) over a 366-day epidemic duration.

Figure 2 A shows the number of unplaced pigs versus time. We observe that the median curve shows a slight increase (less than 1%) after day 130, indicating that the number of pigs not placed in control zones grows very slowly beyond this point for half of the simulations. However, extreme cases continue to increase substantially after day 118, impacting a smaller subset of pigs. Figure 2 B shows that the median percentage of pigs not placed in buffer zones is approximately 55%, slightly higher than in infection zones.

Figure 2, panels C and D display the median number of pigs not placed in control zones across different companies and farm types of origin. Company A carries the overwhelming majority of the unplaced pigs, with over 98% of the total. At the same time, Company B has a small share, and Company C shows a median of zero unplaced pigs. Regarding farm types, most unplaced pigs originate from sow farms 51% and nursery farms 48%, with very few from gilt farms (1%). The median number of unplaced pigs from finisher, farrow-to-finish, and wean-to-finish farms was zero because movements to slaughterhouses were not considered in our analysis (data is unavailable). This pattern aligns with distribution of the pigs transported during the simulation period (Supplementary Material Figure S1).

Figure 3 shows that the medians of infected farms and depopulated pigs begin to stabilize after days 100. This trend indicates that new infections among farms become less frequent after day 100, and the subsequent depopulation also becomes steady in half of the simulations. The point at which the number of unplaced pigs stabilizes occurs at day 130 (Figure 2), 30 days later than depopulation, aligning with the 30-day control duration specified in the NRP scenario. Descriptive statistics of the cumulative number of infected farms and depopulated pigs are presented in Supplementary Material Table S3 and Table S4 respectively.

Figure 4 illustrates the median number of depopulated pigs across different companies and farm types. The cumulative median number of depopulated pigs increases from 28,146 on day 90 to 32,368 on day 366. Most depopulated pigs are from Company A, accounting for 96.5%. Most depopulated pigs were finisher farms (57.9% - 67.6%), followed by nursery farms (24.9% - 25.9%), wean-to-finish farms (0% - 9.3%), and sow farms (6.4% - 7.9%).

3.2. Control zone duration scenarios

Shortening the duration to 20 days (scenario 1) significantly reduced the median number of unplaced pigs (33.5% reduction), with a lower 75th percentile (19.4% reduction) and the maximum (22.6% reduction) compared to the NRP scenario (Table 3). Extending the duration to 25 days (scenario 2) results in the median number of unplaced pigs remaining below the NRP scenario (a 16.7% reduction from the NRP median). Further extending the duration to 40 days (scenario 3) increases the median number of unplaced pigs by 32% compared to the NRP scenario.

The impact of control zone duration on the number of infected farms varies across outbreak severity levels. The median number of infected farms remained unchanged across all scenarios (Table 4). However, shorter control zone duration to 20 and 25 days increased the 75th percentile of infected farms by 18.5% and 5.6%, respectively, suggesting that reducing control duration may lead to higher infection numbers in more severe outbreaks. Conversely, extending the duration to 40 days reduced the 75th percentile by 16.7%, demonstrating that longer control durations are more effective in mitigating the spread of ASF for larger outbreaks.

The median number of depopulated pigs remains broadly consistent across all scenarios (Table 5). However, a higher variation was observed in the 75th percentile among the different scenarios, a control zone of 20 and 25 days showed an increase of 18.7% and 5.4% of depopulated pigs compared to the NRP scenario, respectively, while a control zone of 40 days decreased depopulated pigs by 16.7%.

3.3. Control zone size scenarios

The median number of unplaced pigs increases by 53.6% in scenario 4 (expanded infection zone size 5 km) compared with the NRP scenario (Table 6). In scenario 5 (expanded infection zone size 10 km), it increases substantially by 176.4%, and in scenario 6 (15 km), it increases by 282%.

The median number of infected farms decreased slightly to 6 for scenario 5 (10 km) and scenario 6 (15 km) (Table 7). However, the 75% quantile shows a decrease by 76%. These reductions demonstrate that larger infected zones effectively contain large outbreaks.

Comparing the alternative scenarios with the NRP, our results indicate notable reductions in depopulated pigs (Table 8). In scenario 4 (infected zone of 5 km), the 75th percentile decreases by 37.1%, the median by 3.6%, and the maximum by 10.8%. In scenario 5 (infected zone of 10 km), the 75th percentile dropped by 66%, the median by 12%, and the maximum by 10.5%. Scenario 6 (infected zone of 15 km) achieves the most significant reductions, with the 75th percentile decreasing by 75.7%, the median by 18.4%, and the maximum by 16.2%.

### 3.4. Epidemic elimination rate for each scenario

Under the NRP scenario, where the control zone duration is 30 days and the infection zone radius is 3 km, the elimination rate is 84.5%. Shortening the control duration to 20 days (scenario 1) reduced the elimination rate to 81.9%, while a 25-day duration (scenario 2) to 83.6%. In contrast, extending the control duration to 40 days (scenario 3) increases the elimination rate to 87.9%. Expanding the infection zone radius to 5 km (scenario 4) further increases the elimination rate to 89.8%, while a 10 km radius (scenario 5) results in a significant improvement, reaching 96.9%. The highest elimination rate of 99.1% is observed in scenario 6, where the infection zone radius is extended to 15 km, demonstrating the strongest outbreak containment effect among all tested scenarios.

### 3.5. Cost analysis of depopulated and unplaced pigs

The median price of depopulated pigs remained relatively stable across the NRP and scenarios 1–3, ranging from \$6,001,760 (scenario 3) to \$6,027,840 (scenario 2) (Table 9). However, a noticeable downward trend begins in scenario 4 with a cost of \$5,737,716, and a cost of \$5,071,500, and \$4,535,757 in scenarios 5 and 6 respectively. Conversely, the median price of all unplaced pigs shows an opposite trend, starting at \$4,199,661 under the NRP scenario and decreasing in scenarios 1 and 2 to \$2,783,670 and \$3,488,320, respectively. In scenario 3, unplacement costs rise sharply to \$5,539,086, and the increase becomes more pronounced in scenarios 4–6, peaking at \$16,058,659 in scenario 6.

To evaluate the combined economic impact of depopulation and unplacement, we examined a range of values for $x$, representing the percentage of unplaced pigs requiring culling. We found that when $x = 12\%$, the differences in cost among scenarios (Figure 5 B) were minimized (mathematical proof is shown in Supplementary Material Section Calculation of $x$), indicating that at this threshold, the trade-off between depopulation and unplacement costs is most balanced across all scenarios.

When $x$ was lower than 12% ( $x = 5\%$ in Figure 5 A), the median total cost ranges from $5.3 million (scenario 6 with expanded 15 km infected zone) to $6.2 million (scenario 3 with extended 40 days control zone duration). As $x$ increases to 12% (Figure 5 B), median costs across scenarios converge, ranging from $6.3 million (scenario 1 with shortened 20 day duration) to $6.6 million (scenario 3 with extended 40 days control zone duration). At ( $x$ = 20% in Figure 5 C), scenarios with higher unplacement (scenarios 5 and 6) incur noticeably higher costs, with scenario 6's median rising to $7.8 million, ranking the highest among scenarios. The values of the total cost across each scenario under $x = 5\%$, $x = 12\%$, and $x = 20\%$ are presented in Supplementary Material Table S5.

Across all $x$ values, scenario 6 exhibits the lowest 75th percentile (around $30 million) and maximum (around $610 million) total costs. Conversely, NRP and scenario 1 consistently show the highest maximum (over $800 million) and upper-quartile costs (over $100 million). Notably, in every scenario, the minimum total cost is zero, reflecting simulations where the outbreak is controlled early, with no secondary cases or depopulation.

We also estimated the total cost for each production type when 12% of unplaced pigs were ultimately culled (Figure 6). The results show that the estimated median cost for sow farms is around $4,000,000 for all scenarios, but for scenario 5 (expanded 10 km infected zone) and scenario 6 (expanded 15 km infected zone), their median cost was at $3,284,020 and $2,770,00, respectively. For finisher farms, the estimated median cost remains approximately $1,660,000 under NRP scenario, scenario 1, 2, and 3, with a slight decline to $1,601,968 in scenario 4 (5 km infected zone), $1,473,980 in scenario 5 (10 km infected zone), and $1,362,978 in scenario 6 (15

km infected zone). Nursery farms show a progressive increase in median cost from $405,409 under NRP scenario to $873,242 in scenario 6 (15 km infected zone), with intermediate costs of $334,388 in scenario 1 (20-day duration), $373,824 in scenario 2 (25-day duration), $475,122 in scenario 3 (40-day duration), $511,884 in scenario 4 (5 km infected zone), and $714,980 in scenario 5 (10 km infected zone). For wean-to-finish farms, median costs remain at approximately $270,000 under NRP scenario, scenario 1 (shorter 20-day duration), scenario 2 (moderate 25-day duration), and scenario 3 (longer 40-day duration), but decline sharply to $186,107 in scenario 4 (expanded 5 km infected zone), $100,871 in scenario 5 (expanded 10 km infected zone), and $100,545 in scenario 6 (expanded 15 km infected zone). The median cost for gilt and farrow-to-finish farms are zero.

## 4. Discussion

This study evaluated the impact of control zone movement restrictions on pig placement and economic losses via simulated ASF outbreaks in U.S. commercial swine farms. The model estimated that ASF outbreaks could result in a median of 153,020 (IQR: 27,377 - 1,307,899) unplaced pigs within one year under the current National Response Plan (USDA, 2023). Reducing the duration of control zones to 20 and 25 days decreased the median number of unplaced pigs by 33.5% and 16.7%, respectively. Extending the duration of control zones to 40 days increased the median number of unplaced pigs by 32%. The median number of infected farms and depopulated pigs remained unchanged under all durations. Thus, shorter control zone durations reduced the number of unplaced pigs without exacerbating the epidemic, while extended durations increased unplacement losses without improving ASF control (Halasa et al., 2018). On the other hand, expanding the infection zone radius to 5km, 10km, and 15km reduced the median number of depopulated pigs by 3.6%, 12%, and 18.4%, respectively. However, these benefits came at the cost of significantly higher unplacement, with the median number of unplaced pigs increasing by 53.6% (5 km), 176.4% (10 km), and 282% (15 km). The epidemics elimination rates ranged from 81.9% in scenario 1 (20-day duration, 3 km zone) to 99.1% in scenario 6 (15 km zone expansion), demonstrating a clear trend of increasing elimination rates with longer duration and larger control radii. While all scenarios achieved an elimination rate above 80%, only Scenario 6 successfully eliminated ASF outbreaks in nearly all simulations (>99% elimination rate) within 12 months , highlighting the effectiveness of larger control zones in containing severe outbreaks (Halasa et al., 2018; Sykes et al., 2025). Total economic losses varied across control strategies. Our model found larger control zones were more cost-effective when less than 12% of unplaced pigs were culled. Otherwise, shorter durations and smaller control zones minimized total economic losses.

Simulation of the NRP scenario showed that finisher farms accounted for 58% of depopulated pigs, followed by nursery farms with 25%. In contrast, unplaced pigs were predominantly from nursery farms 48% and sow farms 51%. This result is not surprising given the high level of integration of the U.S. commercial swine industry, with high infection rates at finisher farms expected due to their network position and disproportionate numbers (Angulo et al., 2023; Galvis et al., 2022; Sykes et al., 2023). A larger number of infected farms leads to an increased number of control zones, creating a ripple effect upstream and leading to a backlog at sow and nursery farms. Nursery farms, in particular, face a dual challenge: they must manage the inflow of pigs from sow farms while simultaneously experiencing limited movement options to finisher farms (Cardenas et al., 2024; Rodrigues da Costa et al., 2021). This compounded pressure disrupts normal production cycles and amplifies economic losses due to overcrowding, increased maintenance costs, and potential long-term supply shortages if breeding is adjusted to mitigate unplacement (Tildesley et al., 2019).

For the number of infected farms across all scenarios, one notable change is observed in the 75th percentile of infected farms, from 108 in NRP scenario to 26 farms in scenario 6 by expanding the infection zone radius, while the median number of infected farms remains almost unchanged. This significant shift highlights that more stringent controls are particularly effective in reducing the number of infected farms in moderately severe outbreaks (Sartore et al., 2010). The maximum number of infected farms decreases modestly with stricter control measures, from 1,211 in scenario 1 (shorter 20-day duration) to 948 in scenario 6 (expanded 15 km infected zone), indicating that while severe outbreaks may still occur, their intensity is reduced under stronger control measures. These results suggest that while enlarging control zones may not alter the median epidemic size, it has a substantial impact on controlling more severe outbreaks, reducing both their scale and frequency (Wodarz et al., 2020).

Our results demonstrate that the median and maximum number of unplaced pigs increase as the control zone size expands (Table 6). However, an intriguing observation is the decrease in the 75th percentile of unplaced pigs as the zone size increases. This decrease is likely linked to a corresponding reduction in the 75th percentile of infected farms. As the number of infected farms decreases under stricter control measures, fewer control zones are needed, effectively reducing the total size of the control zones. Thus, while larger control zones inherently increase the number of unplaced pigs in most cases, their ability to significantly reduce infections and limit the creation of new control zones may mitigate their overall impact on unplacement, particularly in outbreaks with a median of 108 infected farms. Previous research by Halasa et al.(Halasa et al., 2018) focused on the economic and epidemiological effects of movement restrictions and control zone size, concluding that larger control zones tend to increase economic burden without substantial epidemiological benefits. Our study aligns with their observations but adds a new dimension by demonstrating that larger zones can still be beneficial in reducing both unplacement and depopulation of severe outbreaks.

We have demonstrated that the percentage of unplaced pigs that are culled influences the estimated total economic cost. Our analysis revealed a critical threshold where the optimal control strategy minimizes the total economic losses shifts. When less than 12% of unplaced pigs are culled, scenarios with larger control zones are less costly because the reduction in depopulation costs outweighs the relatively minor unplacement losses. Conversely, when more than 12% of unplaced pigs require culling, scenarios with shorter durations and smaller control zones become more cost-effective, as the economic burden of unplaced pigs dominates the total cost. The 12% threshold identified in this study represents the cutoff point at which the total costs across different control strategies are most consistent and closely aligned, as it minimizes cost variability. However, this value is scenario-dependent; if key parameters of control strategies such as control zone size or duration were altered, the threshold would likely shift. We emphasize the 12% threshold in this study because, within our simulation framework, it marks a distinct change in how total costs respond to varying ASF control strategies.

Among the scenarios, scenario 6 (expanded 15 km infection zone) consistently demonstrates the lowest 75th percentile and maximum total costs across all values of $x$, emphasizing its ability to limit extreme economic losses during outbreaks (Figure 5). In contrast, the NRP scenario and scenario 1 regularly incur the highest upper-quartile and maximum costs, primarily due to their short control durations and smaller infected zone sizes, which result in elevated farm infections and depopulation rates in extreme cases. Additionally, Scenario 6 maintains the smallest interquartile range (IQR) for each *x* value, indicating reduced variability and lower financial risk.

**5. Limitations and final remarks**

This study estimated the number of unplaced pigs due to movement restrictions in ASF control zones. However, the ability to move pigs out of these zones is contingent on obtaining movement permits (USDA, 2023), which can lead to delays, due to testing and administrative constraints. As demonstrated by Sykes et al. (Sykes et al., 2023), during the first 60 days of outbreaks, a median of 1808 (IQR: 0–9737) permits were required. While awaiting approvals, farms within control zones continue to incur feed, housing, and management costs, adding to the economic burden. Future research should incorporate the full spectrum of movement restriction dynamics, including incoming, outgoing,  and internal movements of pigs and vehicles, and personnel interactions within control zones. A more comprehensive understanding of these movements will provide a clearer picture of the economic and logistical challenges of ASF-related restrictions.

Another limitation was the lack of slaughterhouse movements (Cardenas et al., 2024). This particularly affects the analysis of unplacement at finisher farms, wean-to-finish farms, and farrow-to-finish farms, which mostly send pigs to slaughter. Holding market-ready pigs at finisher farms for extended periods can also lead to economic losses due to increased maintenance costs. By not including the movements to slaughterhouses, the study may have also underestimated the financial impacts of ASF-related restrictions on growing pig farms. Incorporating movement data to slaughterhouses, as it becomes available, in future studies would enhance the approximation of economic impact.

Additionally, pig prices vary over time. The market value of pigs fluctuates due to factors such as seasonal demand, feed costs, and market supply dynamics. Disruptions in the supply chain could cause price volatility, altering the relative economic impact of unplacement and depopulation costs.

At last, our analysis only accounts for the economic losses due to the culling of unplaced pigs, without considering the additional costs incurred to keep pigs for an extended period, such as additional feed, personnel, pig market value, and culling cost. Future research should incorporate these itemized costs to provide a more comprehensive assessment of the economic burden of the unplacement of pigs.

**6. Conclusion**

We evaluated the impact of ASF control zone movement restrictions on pig placement and economic losses, highlighting the trade-offs between disease containment and industry-wide disruptions. Our findings show that shorter control durations effectively reduce the burden of unplaced pigs without compromising outbreak control, whereas prolonged restrictions exacerbate economic losses with limited epidemiological benefits. Similarly, expanding control zones enhances containment of severe outbreaks but significantly increases unplacement, underscoring the need for balanced strategies. Our economic analysis revealed that the proportion of unplaced pigs requiring culling ($x$) is key to determining the optimal strategy for the median total costs. When fewer than 12% of unplaced pigs were culled, larger control zones (scenario 6) minimized median total costs by reducing depopulation. However, when more than 12% were culled, shorter durations and smaller zones (scenario 1) proved more cost-effective by limiting unplacement. These results emphasize the need for region-specific policies that balance depopulation and unplacement costs. Targeted resource allocation, such as strategies for processing or rerouting unplaced pigs, can help mitigate economic burdens and support stricter containment measures without excessive financial strain. Adaptive ASF control strategies, informed by outbreak severity and cost management, could improve the swine industry's resilience to future outbreaks.

**Funding**

This project is funded by National Institute of Food and Agriculture; Food and Agriculture Cyber informatics and Tools, Grant/Award Numbers:2020-67021-32462 and by the Foundation for Food Agriculture Research (FFAR) award number FF-NIA21-0000000064.

**Ethical statement**

The authors confirm the ethical policies of the journal, as noted on the journal's author guidelines page. Since this work did not involve animal sampling nor questionnaire data collection by the researchers, there was no need for ethics permits.

**Data Availability**

The data that support the findings of this study are not publicly available and are protected by confidential agreements. More information about the model implementation can be found at https://github.com/machado-lab/PigSpread- ASF.

**Acknowledgments**

The authors would like to acknowledge participating companies and veterinarians for their insists and concerns regarding the impact of unplacement of pigs due to control zones.

**List of tables**

**Table 1:** Control zone duration and infected zone radius across seven scenarios

| **Scenario** | **NRP*** | **1** | **2** | **3** | **4** | **5** | **6** |
|---|---|---|---|---|---|---|---|
| **Duration (days)** | 30 | 20 | 25 | 40 | 30 | 30 | 30 |
| **Radius of IZ (km)** | 3 | 3 | 3 | 3 | 5 | 10 | 15 |

*NRP = National Response Plan (USDA, 2023).

**Table 2:** Market average price for a pig from farm type ($i$) (Eadie, 2024; NutriQuest, 2024)

| **Farm type ($i$)** | Sow | Nursery | Finisher | Gilt | Wean-to-finish | Farrow-to-finish |
|---|---|---|---|---|---|---|
| **Depopulation price ($p_i^d$)** | $1,449 | $25.97 | $92.74 | $350 | $92.74 | $92.74 |
| **Unplacement price ($p_i^u$)** | $25.97 | $25.97 | $92.74 | $350 | $92.74 | $92.74 |

**Table 3.** The cumulative number of pigs not placed into control zones for the National Response Plan (NRP) and the alternative control scenarios over a 366-day epidemic duration

| **Control zone duration** | **Minimum** | **25% quantile** | **Median** | **75% quantile** | **Maximum** |
|---|---|---|---|---|---|
| 30 days (NRP) | 0 | 27,377 | 153,020 | 1,307,899 | 4,556,749 |
| 20 days | 0 | 16,272 | 101,718 | 1,051,884 | 3,528,501 |
| 25 days | 0 | 22,012 | 127,506 | 1,161,401 | 4,256,250 |
| 40 days | 0 | 38,686 | 201,969 | 1,487,950 | 5,235,828 |

**Table 4.** The cumulative number of infected farms for the National Response Plan (NRP) and the alternative control scenarios over a 366-day epidemic duration

| Control zone duration | Minimum | 25% quantile | Median | 75% quantile | Maximum |
|---|---|---|---|---|---|
| 30 days (NRP) | 0 | 1 | 7 | 108 | 1,192 |
| 20 days | 0 | 1 | 7 | 128 | 1,211 |
| 25 days | 0 | 1 | 7 | 114 | 1,173 |
| 40 days | 0 | 1 | 7 | 90 | 1,129 |

**Table 5.** The cumulative number of depopulated pigs for the National Response Plan (NRP) and the alternative control scenarios over a 366-day epidemic duration

| Control zone duration | Minimum | 25% quantile | Median | 75% quantile | Maximum |
|---|---|---|---|---|---|
| 30 days (NRP) | 0 | 6,480 | 32,368 | 499,495 | 3,940,829 |
| 20 days | 0 | 6,480 | 32,167 | 592,926 | 3,996,183 |
| 25 days | 0 | 6,475 | 32,350 | 526,377 | 4,030,652 |
| 40 days | 0 | 6,420 | 32,198 | 419,241 | 3,868,155 |

**Table 6.** The cumulative number of unplaced pigs for the National Response Plan (NRP) and the alternative control scenarios over a 366-day epidemic duration

| Infection zone radius | Minimum | 25% quantile | Median | 75% quantile | Maximum |
|---|---|---|---|---|---|
| 3km (NRP) | 0 | 27,377 | 153,020 | 1,307,899 | 4,556,749 |
| 5 km | 0 | 50,975 | 235,062 | 1,200,460 | 4,651,382 |
| 10 km | 0 | 131,739 | 422,884 | 1,135,710 | 5,391,554 |
| 15 km | 0 | 220,515 | 584,608 | 1,171,930 | 6,207,919 |

**Table 7.** The cumulative number of infected farms for the National Response Plan (NRP) and the alternative control scenarios over a 366-day epidemic duration

| Infection zone radius | Minimum | 25% quantile | Median | 75% quantile | Maximum |
|---|---|---|---|---|---|
| 3km (NRP) | 0 | 1 | 7 | 108 | 1,192 |
| 5 km | 0 | 1 | 7 | 68 | 1,039 |
| 10 km | 0 | 1 | 6 | 37 | 999 |
| 15 km | 0 | 1 | 6 | 26 | 948 |

**Table 8.** The cumulative number of depopulated pigs for the National Response Plan (NRP) and the alternative control scenarios over a 366-day epidemic duration

| Infection zone radius | Minimum | 25% quantile | Median | 75% quantile | Maximum |
|---|---|---|---|---|---|
| 3km (NRP) | 0 | 6,480 | 32,368 | 499,495 | 3,940,829 |
| 5 km | 0 | 6,472 | 31,200 | 314,280 | 3,517,405 |
| 10 km | 0 | 6,480 | 28,506 | 169,996 | 3,528,729 |
| 15 km | 0 | 6,480 | 26,410 | 121,493 | 3,303,188 |

**Table 9.** Estimated median financial impacts due to depopulation and unplacement

| Scenario | NRP | 1 | 2 | 3 | 4 | 5 | 6 |
|---|---|---|---|---|---|---|---|
| **Depopulation** | | | | | | | |
| **25% quantile** | $662,434 | $653,539 | $656,599 | $652,890 | $655,875 | $652,890 | $652,890 |
| **Median** | $6,008,517 | $6,005,450 | $6,027,840 | $6,001,760 | $5,737,716 | $5,071,500 | $4,535,757 |
| **75% quantile** | $96,311,538 | $113,921,920 | $102,301,608 | $81,345,332 | $61,972,275 | $33,573,532 | $23,339,458 |
| **Unplacement** | | | | | | | |
| **25% quantile** | $785,021 | $464,655 | $625,893 | $1,103,712 | $1,473,573 | $3,807,105 | $6,324,186 |
| **Median** | $4,199,661 | $2,783,670 | $3,488,320 | $5,539,086 | $6,417,321 | $11,580,382 | $16,058,659 |
| **75% quantile** | $35,631,937 | $28,569,403 | $31,642,072 | $40,607,957 | $32,859,479 | $31,465,112 | $32,726,137 |

*In the National Response Plan scenario (**NRP**), control zone duration is 30 days, infection zone radius is 3 km; in **scenario 1**, control zone duration is 20 days, infection zone radius is 3 km; in **scenario 2**, control zone duration is 25 days, infection zone radius is 3 km; in **scenario 3**, control zone duration is 40 days, infection zone radius is 3 km; in **scenario 4**, control zone duration is 30 days, infection zone radius is 5 km; in **scenario 5**, control zone duration is 30 days, infection zone radius is 10 km; in **scenario 6**, control zone duration is 30 days, infection zone radius is 15 km.

**Figures legend**

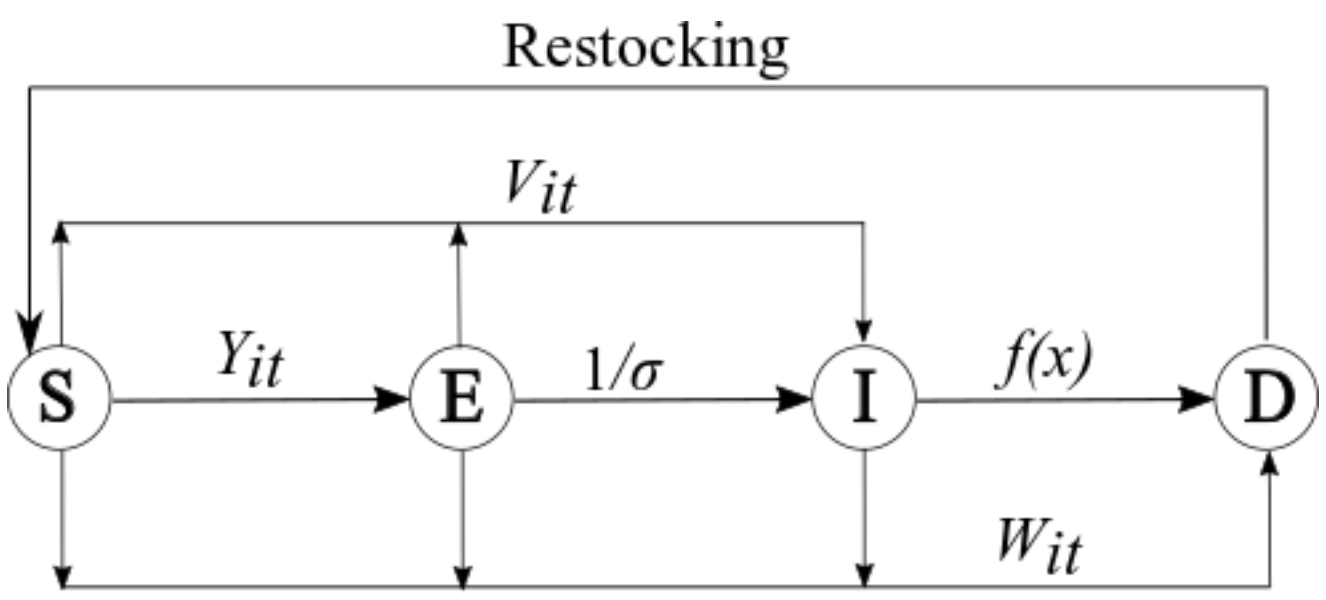

**Figure 1. Schematic diagram of the transitions between different health compartments in the SEID transmission model.** The terms $Y_{it}$, $V_{it}$, and $W_{it}$ represent the probability of transitioning into the exposed (E), infected (I), or detected (D) compartments, respectively, based on the receipt of an exposed, infected, or detected swine movement. The latent period of the ASF virus is denoted by $\sigma$. The equation $f(x)$ represents detection rate.

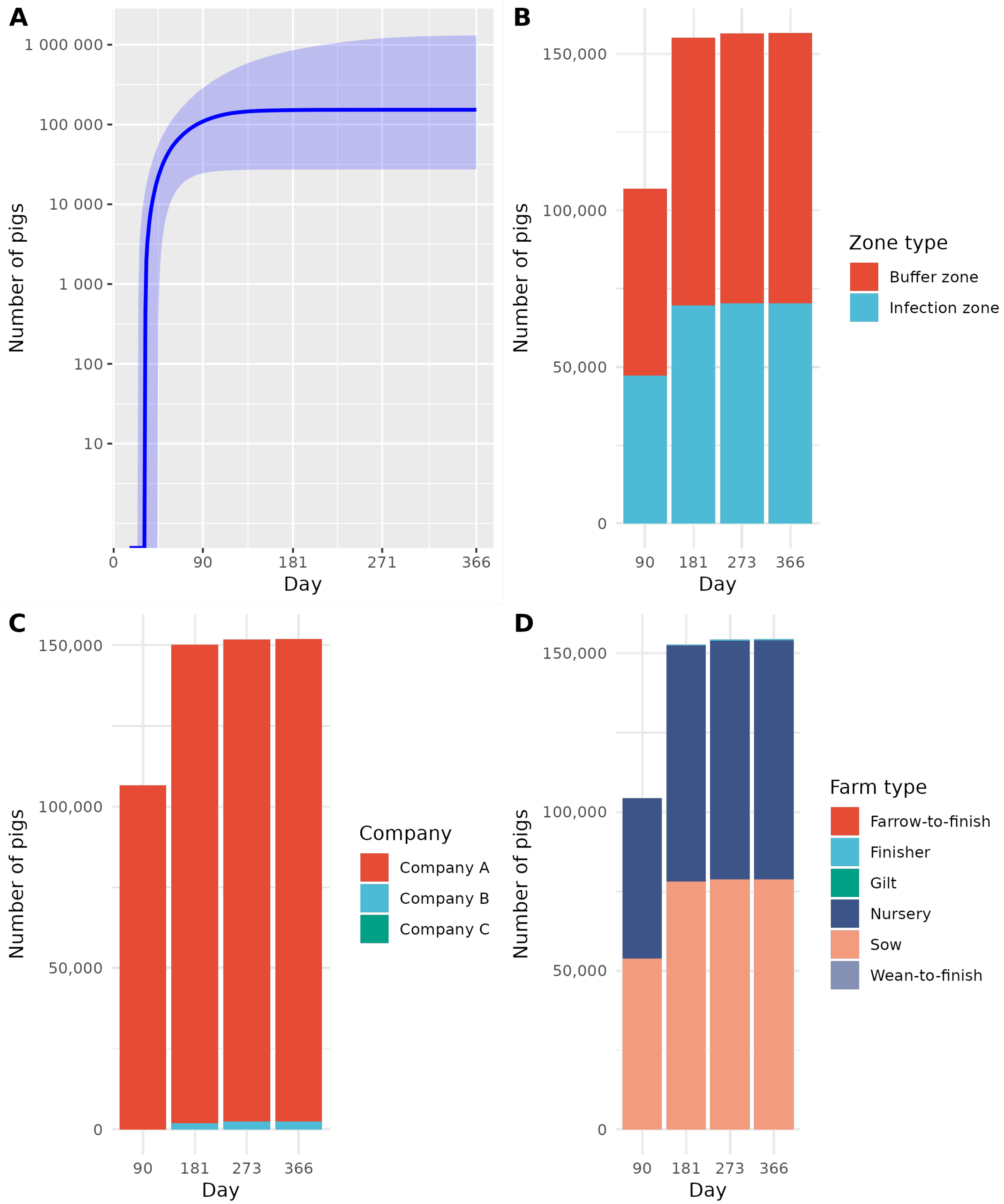

**Figure 2. Cumulative number of unplaced pigs**. A) The daily number of pigs not placed for 366 days, where the solid curve represents the median value, and the shaded area represents the 25% quantile to 75% quantile; B) the median number of pigs not placed categorized by infection

and buffer zone type; C) the median number of pigs not placed within control zones categorized by companies; D) the median number of pigs not placed within control zones categorized by origin farm type.

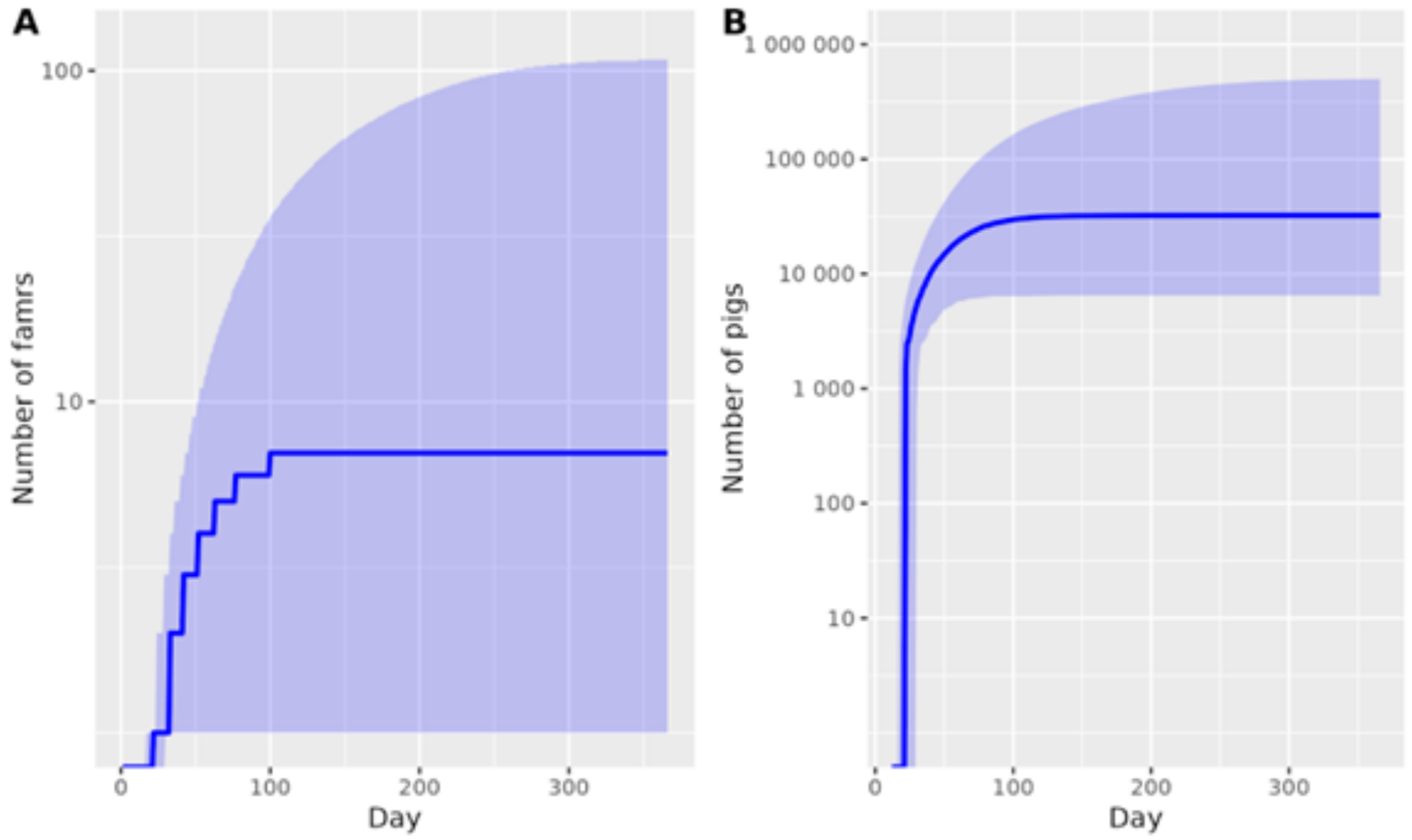

**Figure 3. Cumulative number of infected farms and depopulated pigs for the National Response Plan (NRP) scenario.** A) The daily number of infected farms for 366 days; B) The daily number of depopulated pigs for 366 days, where the solid curve represents the median value, and the shaded area represents the 25% quantile to 75% quantile.

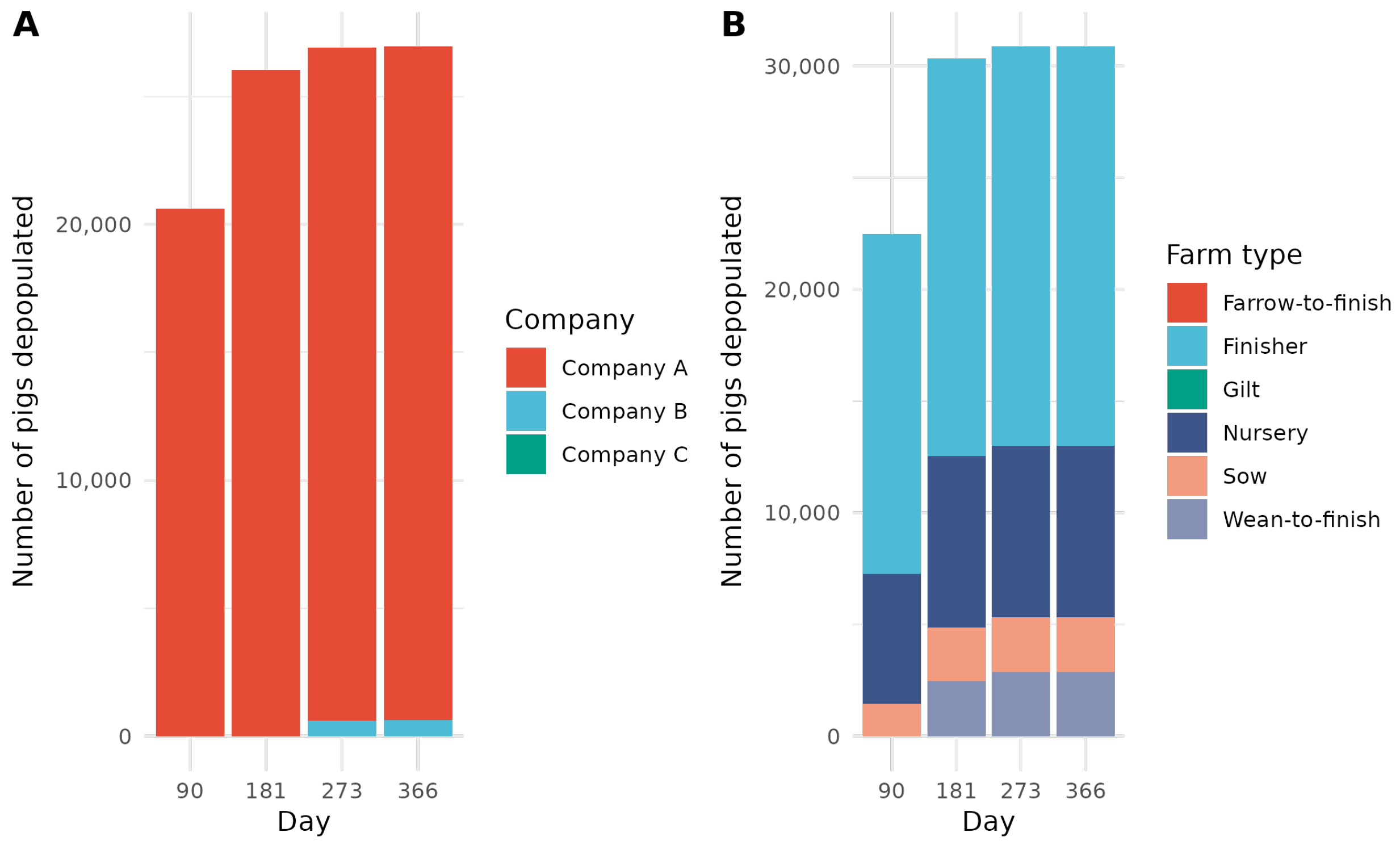

**Figure 4. Median number of depopulated pigs.** A) categorized by companies; B) by farm types.

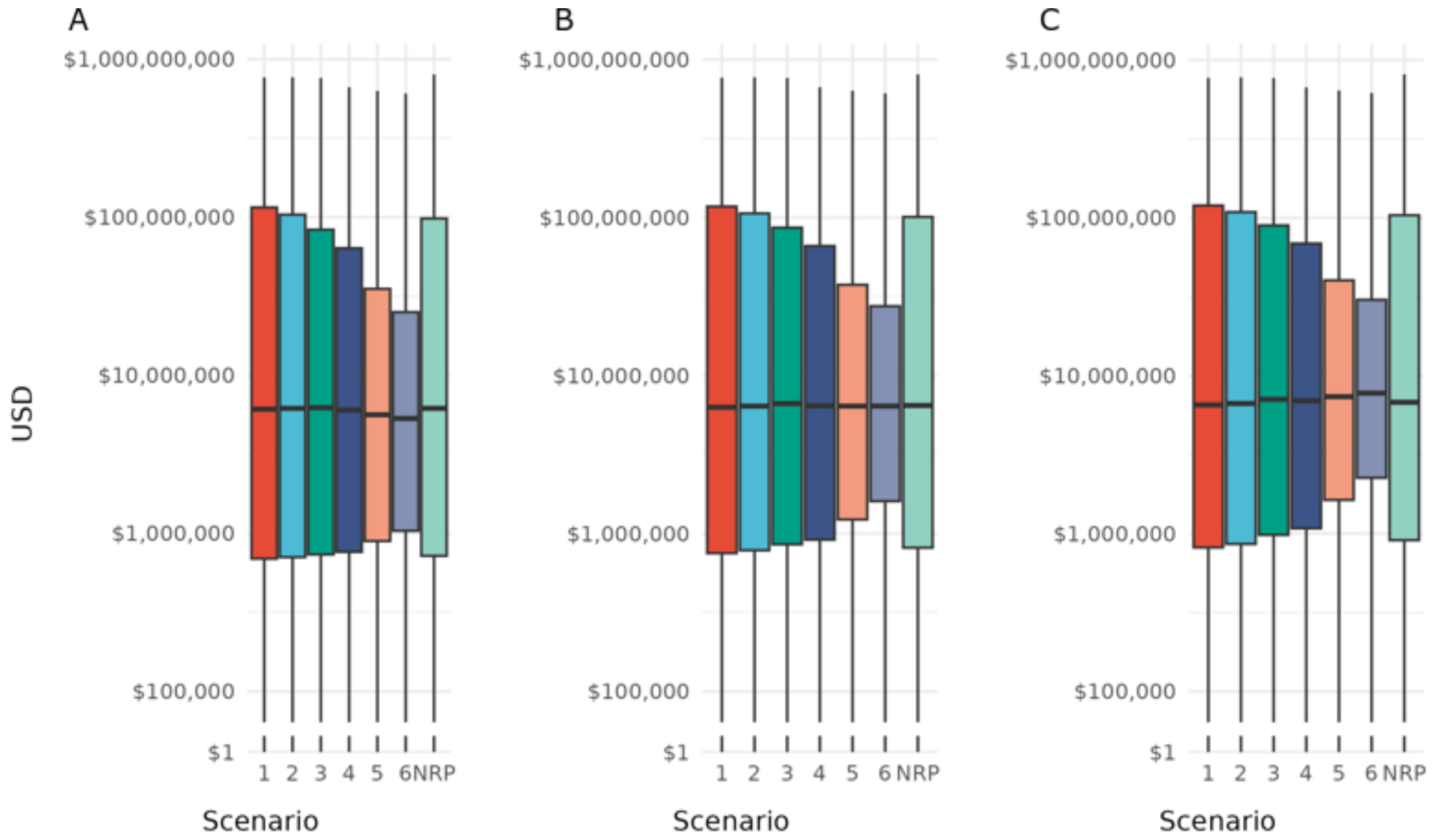

**Figure 5. Distribution of estimated economic cost $L = D + x \cdot U$ across different scenarios.** A) Scenario in which 5% of unplaced pigs were culled; B) Scenario with 12% of unplaced pigs were culled; C) Scenario in which 20% of unplaced pigs were culled, where y-axis is displayed using

a logarithmic scale with base 10. In the National Response Plan scenario (NRP), control zone duration is 30 days, infection zone radius is 3 km; in scenario 1, control zone duration is 20 days, infection zone radius is 3 km; in scenario 2, control zone duration is 25 days, infection zone radius is 3 km; in scenario 3, control zone duration is 40 days, infection zone radius is 3 km; in scenario 4, control zone duration is 30 days, infection zone radius is 5 km; in scenario 5, control zone duration is 30 days, infection zone radius is 10 km; in scenario 6, control zone duration is 30 days, infection zone radius is 15 km.

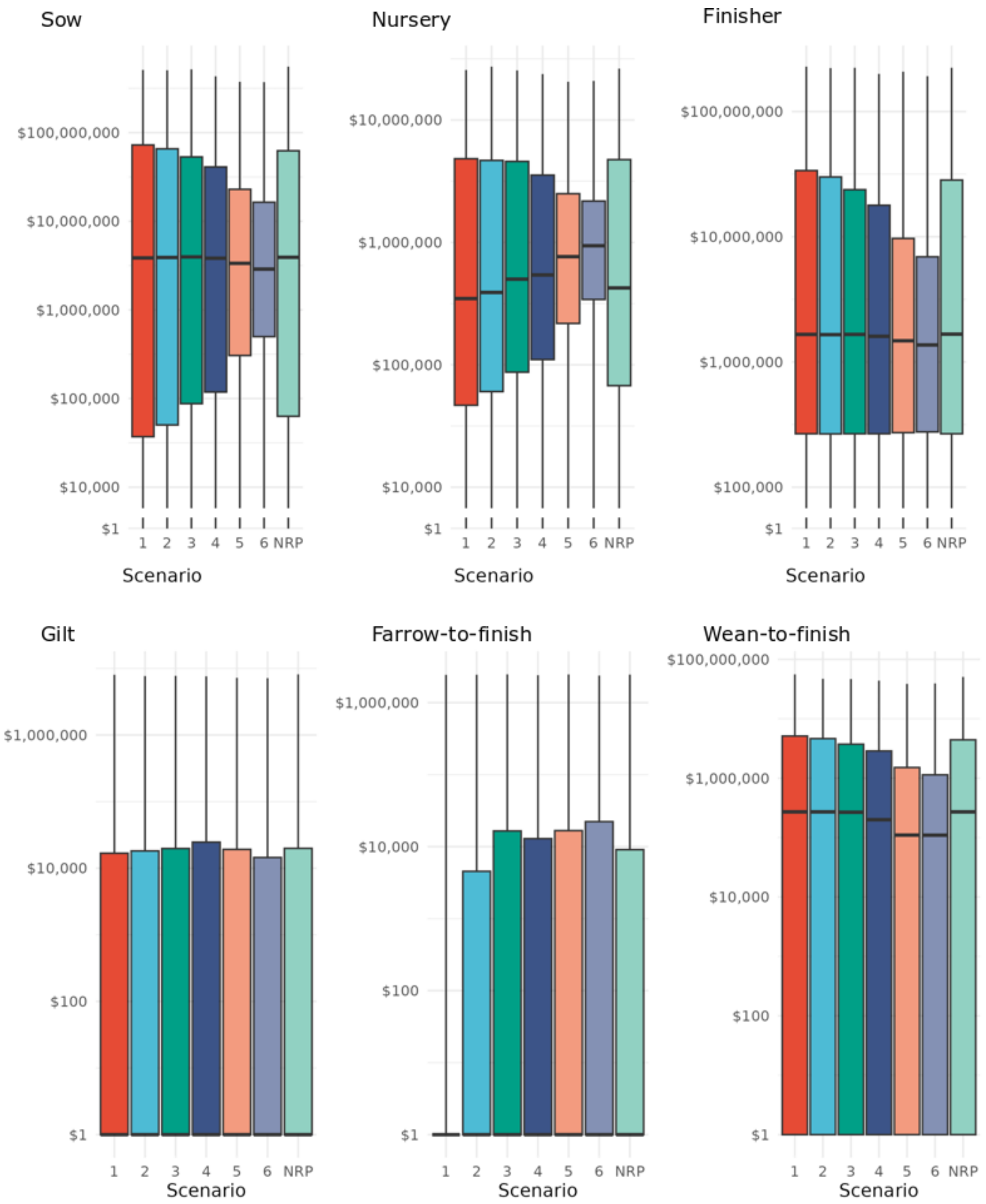

**Figure 6. Distribution of estimated economic cost for each production type across different scenarios when 12% of unplaced pigs were culled.** The y-axis is displayed using a logarithmic scale with base 10. In the National Response Plan scenario (NRP), control zone duration is 30 days, infection zone radius is 3 km; in scenario 1, control zone duration is 20 days, infection zone radius is 3 km; in scenario 2, control zone duration is 25 days, infection zone radius is 3 km; in scenario 3, control zone duration is 40 days, infection zone radius is 3 km; in scenario 4, control zone duration is 30 days,

infection zone radius is 5 km; in scenario 5, control zone duration is 30 days, infection zone radius is 10 km; in scenario 6, control zone duration is 30 days, infection zone radius is 15 km.

**Modeling the impact of control zone restrictions on pig placement in simulated African swine fever in the United States**

**Table S1.** Parameter values

| **Parameter** | **Notation** | **Median particle value (IQR)** |
|---|---|---|
| Transmission rate for exposed pig movements | $\beta_e$ | 1.51 (1.49 - 1.52) |
| Transmission rate for infected pig movements | $\beta_i$ | 1.51 (1.50 - 1.52) |
| Transmission rate for detected pig movements | $\beta_d$ | 1.51 (1.50 - 1.52) |
| Transmission rate for movements of pig trucks | $\beta_p$ | 0.0614 (0.0594 - 0.0627) |
| Transmission rate for movements of market trucks | $\beta_m$ | 0.0606 (0.0576 - 0.0628) |
| Transmission rate for feed delivery trucks | $\beta_f$ | 0.0288 (0.0283 - 0.0293) |
| Transmission rate for other vehicles | $\beta_o$ | 0.0605 (0.0589 - 0.0621) |
| Maximum probability of local transmission | $\phi$ | 0.0630 (0.0620 - 0.0643) |
| The gradient of transmission probability decay | $\alpha$ | 1.39 (1.39 - 1.39) |
| Effective surveillance in sow farms | $L_{sow}$ | 0.851 (0.769 - 0.929) |
| Effective surveillance in nursery farms | $L_{nursery}$ | 0.634 (0.586 - 0.689) |
| Effective surveillance in finisher farms | $L_{finisher}$ | 0.556 (0.544 - 0.576) |

**Table S2.** Descriptive statistics of the cumulative number of unplaced pigs in control zones for the National Response Plan (NRP) scenario

| **Day** | **Minimum** | **25% quantile** | **Median** | **75% quantile** | **Maximum** |
|---|---|---|---|---|---|
| 90 | 0 | 24,468 | 109,149 | 286,088 | 1,317,000 |
| 181 | 0 | 27,296 | 151,676 | 851,929 | 2,295,878 |
| 273 | 0 | 27,372 | 152,953 | 1,209,067 | 3,407,798 |
| 366 | 0 | 27,377 | 153,020 | 1,307,899 | 4,556,749 |

**Pig movement data**

The total number of pigs transported between farms during the simulation period was 83,556,303. Among these, 98.06% were moved by Company A, 1.93% by Company B, and 0.01 % by Company C. 54.03% of the pigs were transported from sow farms, followed by 42.15% from nursery farms, 1.26% from wean-to-finish farms, and the remainder from farrow-to-finish, finisher, and gilt farms(Figure S2).

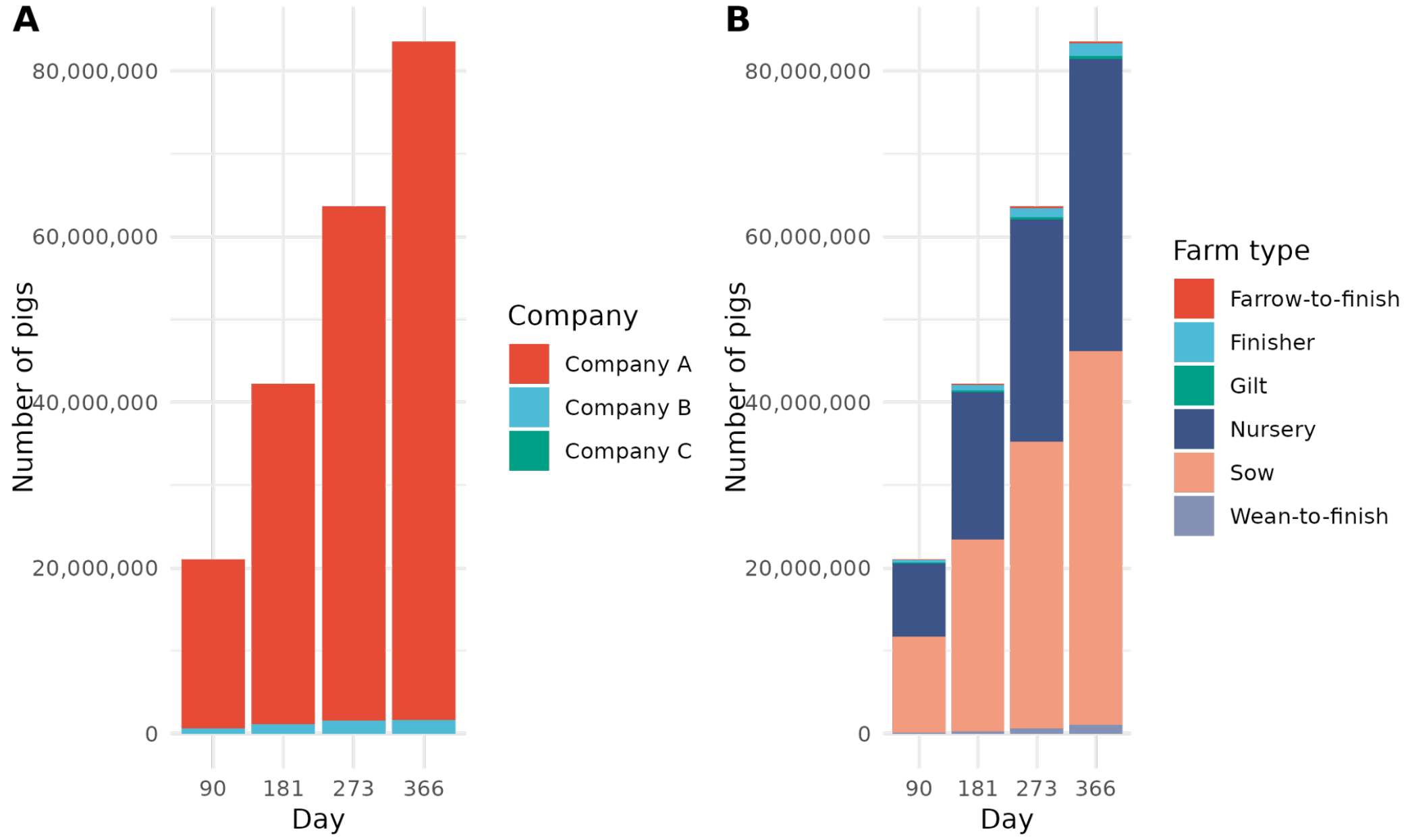

**Figure S1. Cumulative number of pigs transported between farms at 90, 181, 273, and 366 days.** A) Barplot is categorized by source company, and B) by source farm type.

**Table S3**. Descriptive statistics of the cumulative number of infected farms for the National Response Plan (NRP) scenario

| **Day** | **Minimum** | **25% quantile** | **Median** | **75% quantile** | **Maximum** |
|---|---|---|---|---|---|
| 90 | 1 | 1 | 6 | 30 | 346 |
| 181 | 1 | 1 | 7 | 76 | 743 |
| 273 | 1 | 1 | 7 | 102 | 991 |
| 366 | 1 | 1 | 7 | 108 | 1,192 |

**Table S4**. Descriptive statistics of the cumulative number of depopulated pigs for the National Response Plan (NRP) scenario

| Day | Minimum | 25% quantile | Median | 75% quantile | Maximum |
|---|---|---|---|---|---|
| 90 | 0 | 6,400 | 28,146 | 136,848 | 1,168,562 |
| 181 | 0 | 6,480 | 32,261 | 348,743 | 2,399,759 |
| 273 | 0 | 6,480 | 32,360 | 470,173 | 3,255,218 |
| 366 | 0 | 6,480 | 32,368 | 499,495 | 3,940,829 |

**Scenario 1** involves a shortened control zone duration of 20 days, compared to the 30-day duration specified in the NRP (National Response Plan). In this scenario, all other control measures remain consistent with the NRP.

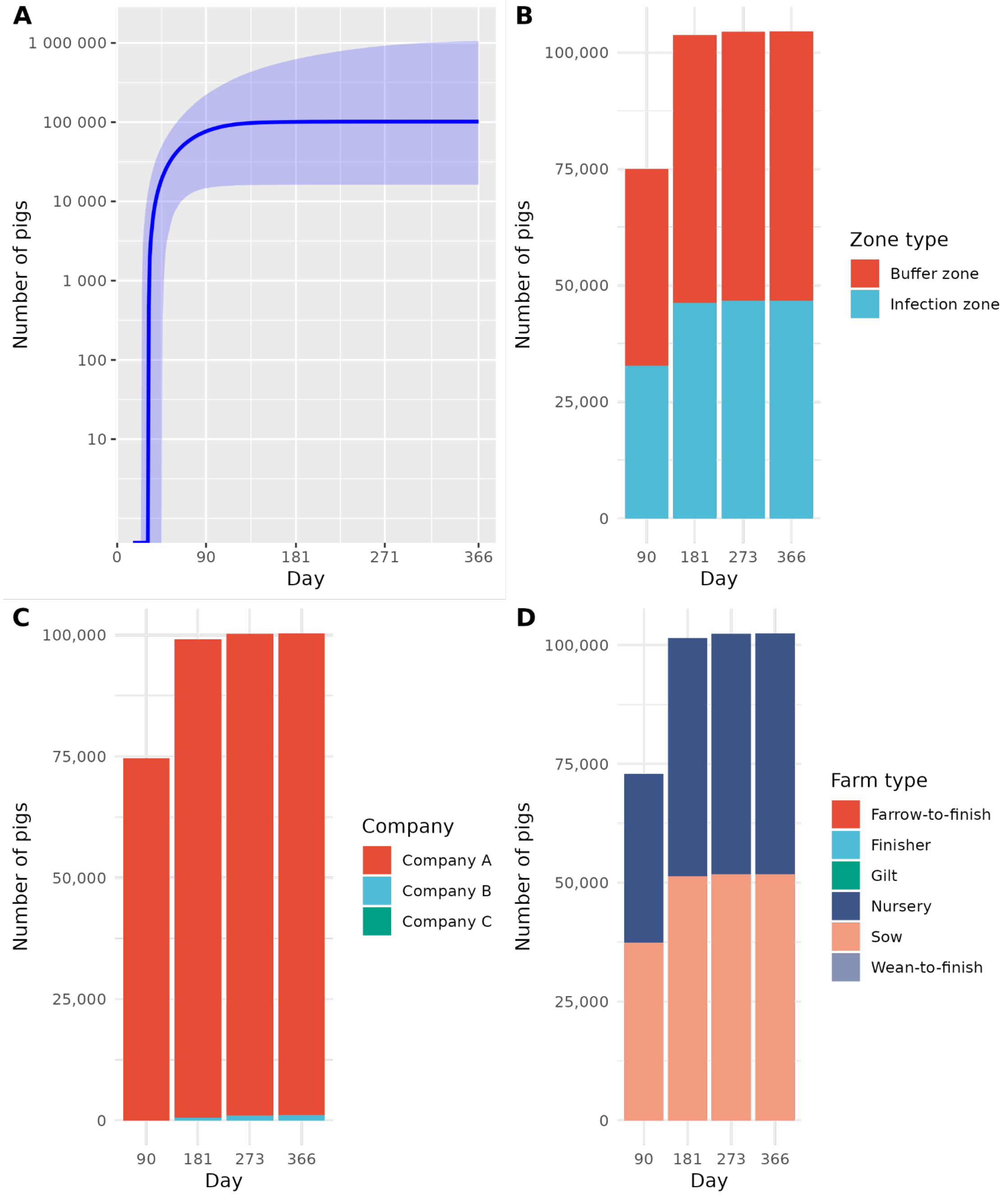

**Figure S2. Cumulative number of unplaced pigs for scenario 1**. A) The daily number of pigs not placed transformed on a logarithmic scale for 366 days, where the solid curve represents the median value, and the shaded area represents the 25% quantile to 75% quantile; B) the median number of pigs not placed categorized by infection and buffer zone type; C) the median number of pigs not placed within control zones categorized by companies; D) the median number of pigs not placed within control zones categorized by origin farm type.

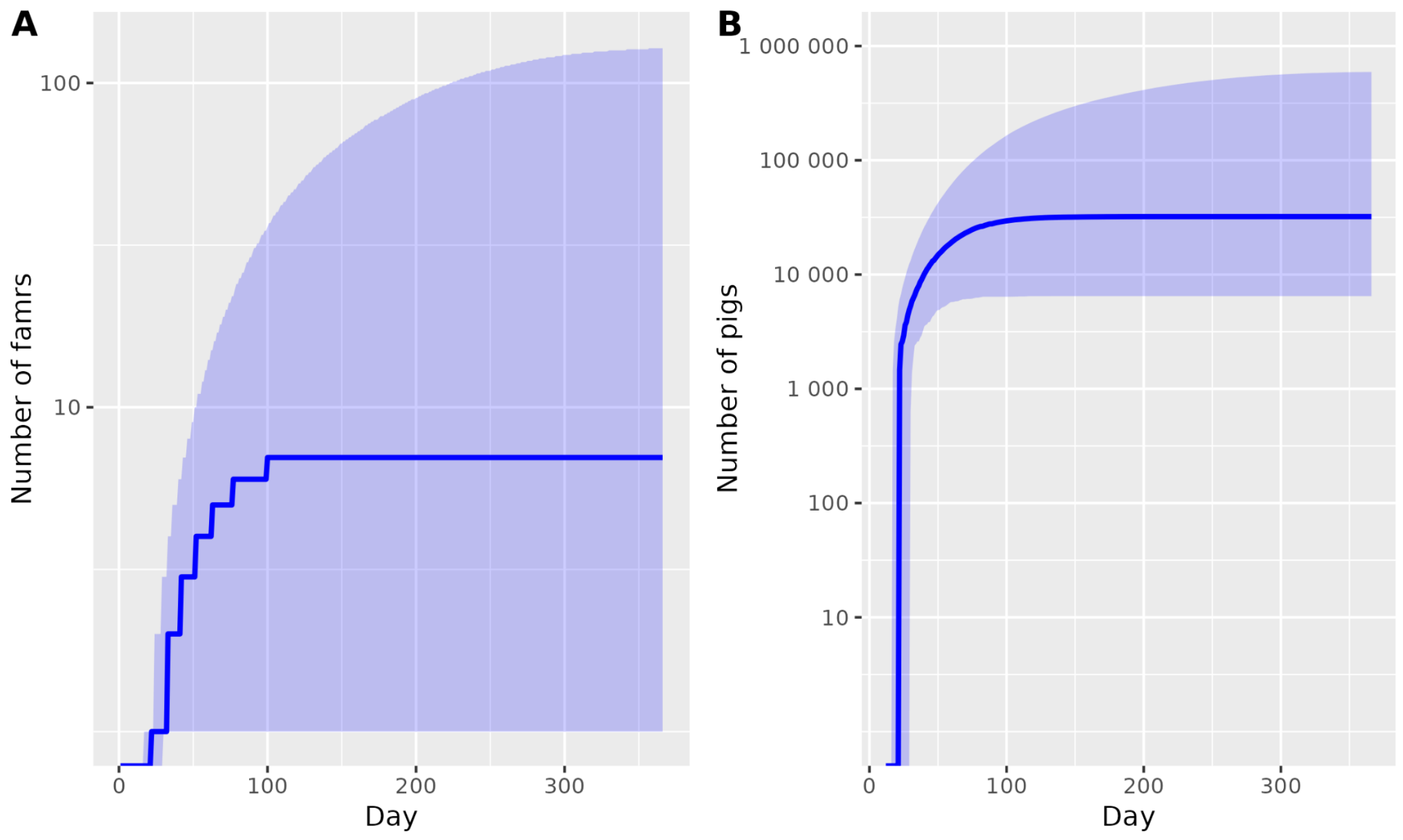

**Figure S3. Cumulative number of infected farms and depopulated pigs for scenario 1.** A) The daily number of infected farms transformed on a logarithmic scale for 366 days; B) The daily number of depopulated pigs transformed on a logarithmic scale for 366 days, where the solid curve represents the median value, and the shaded area represents the 25% quantile to 75% quantile.

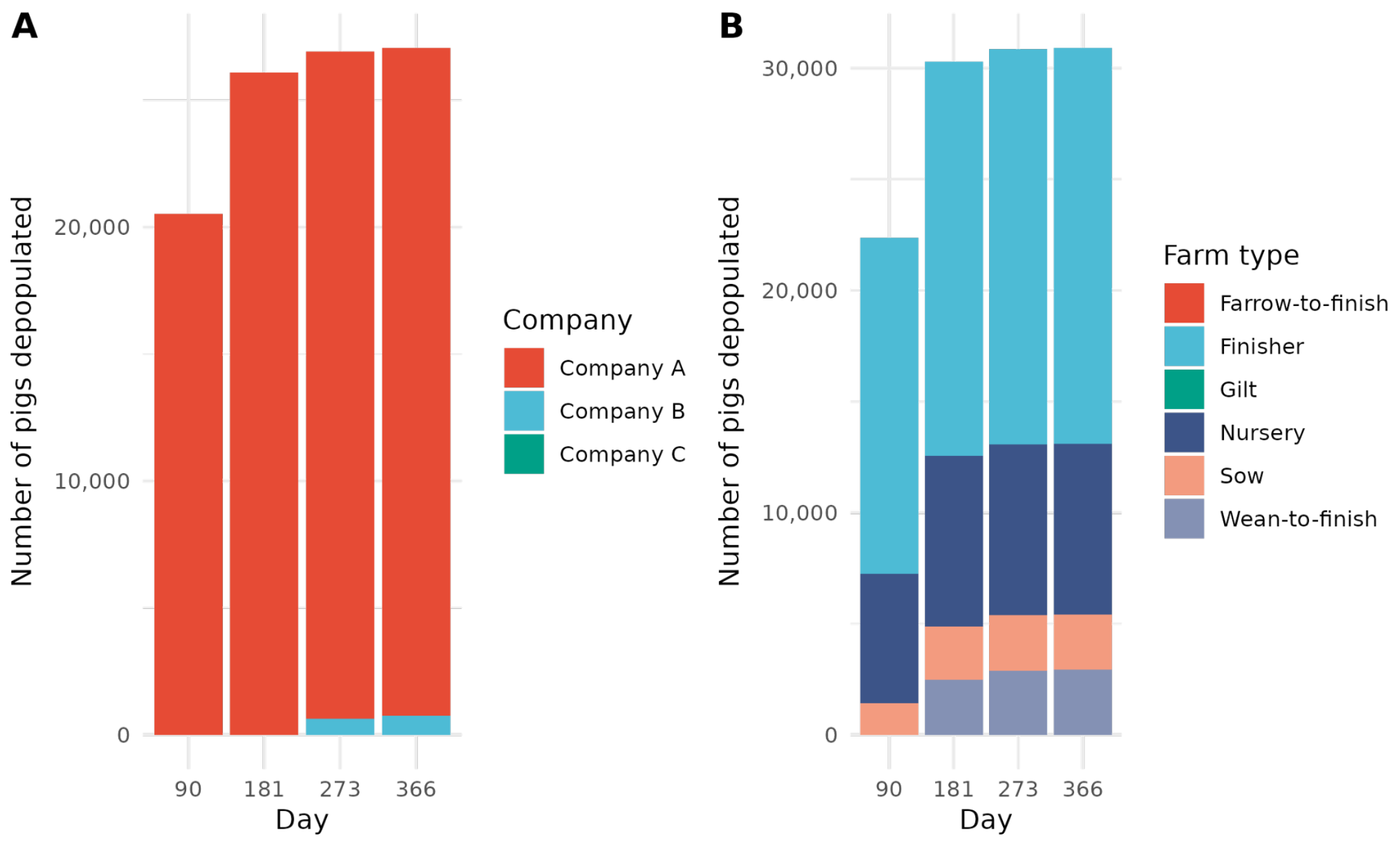

**Figure S4. Median number of depopulated pigs for scenario 1.** A) categorized by companies; B) by farm types.

**Scenario 2** involves a shortened control zone duration of 25 days, compared to the 30-day duration specified in the NRP (National Response Plan). In this scenario, all other control measures remain consistent with the NRP.

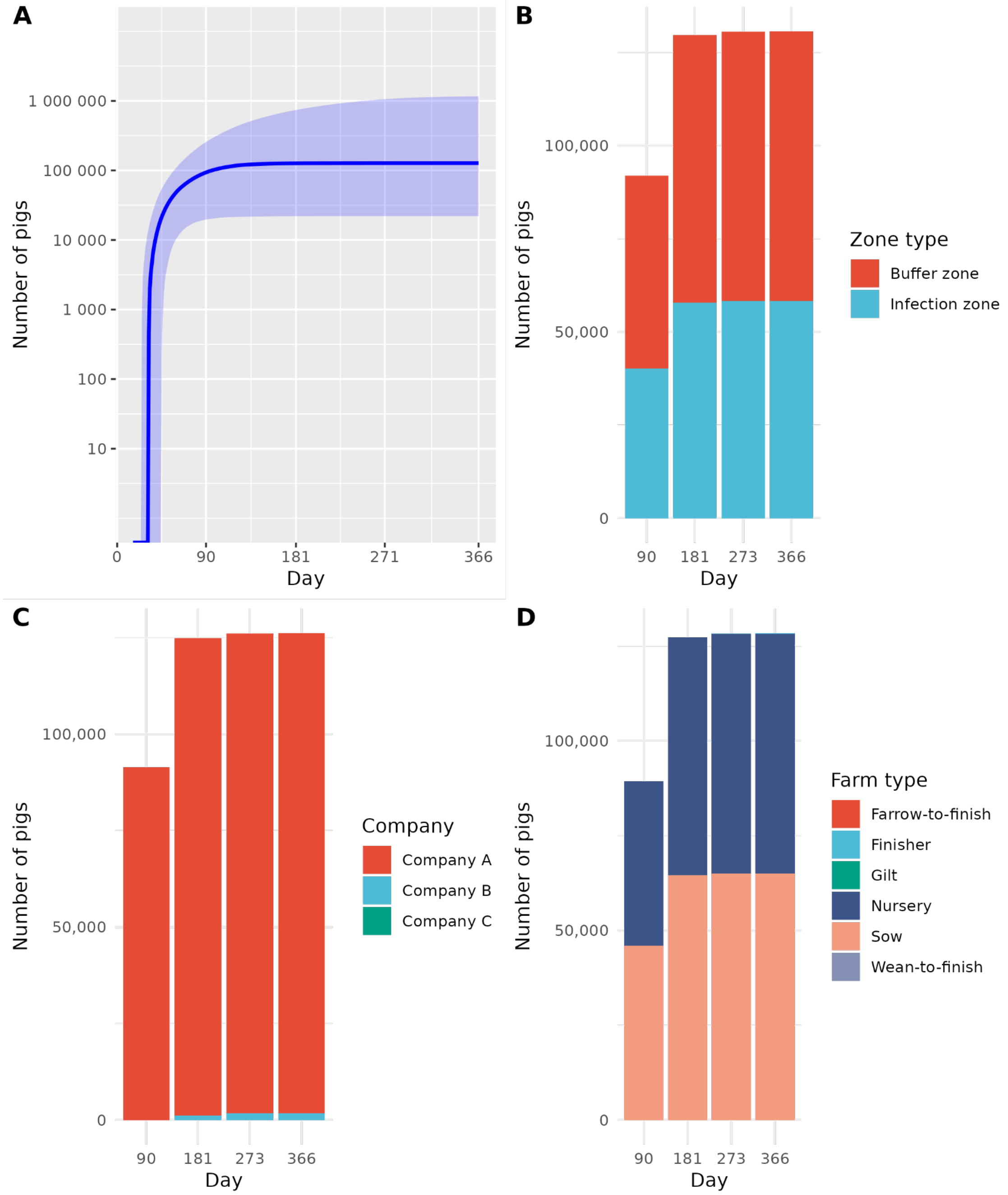

**Figure S5. Cumulative number of unplaced pigs for scenario 2**. A) The daily number of pigs not placed transformed on a logarithmic scale for 366 days, where the solid curve represents the median value, and the shaded area represents the 25% quantile to 75% quantile; B) the median number of pigs not placed categorized by infection and buffer zone type; C) the median number of pigs not placed within control zones categorized by companies; D) the median number of pigs not placed within control zones categorized by origin farm type.

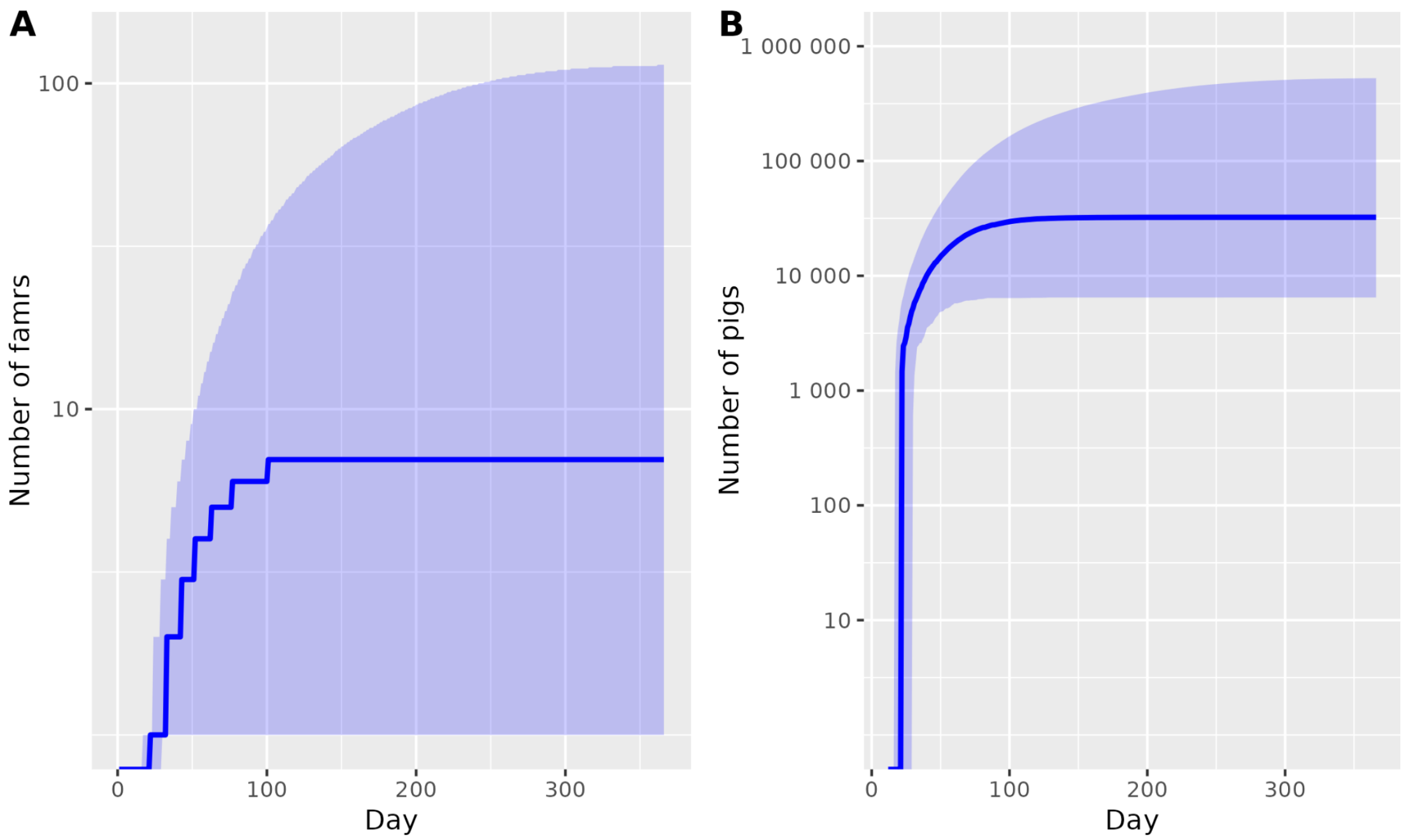

**Figure S6. Cumulative number of infected farms and depopulated pigs for scenario 2.** A) The daily number of infected farms transformed on a logarithmic scale for 366 days; B) The daily number of depopulated pigs transformed on a logarithmic scale for 366 days, where the solid curve represents the median value, and the shaded area represents the 25% quantile to 75% quantile.

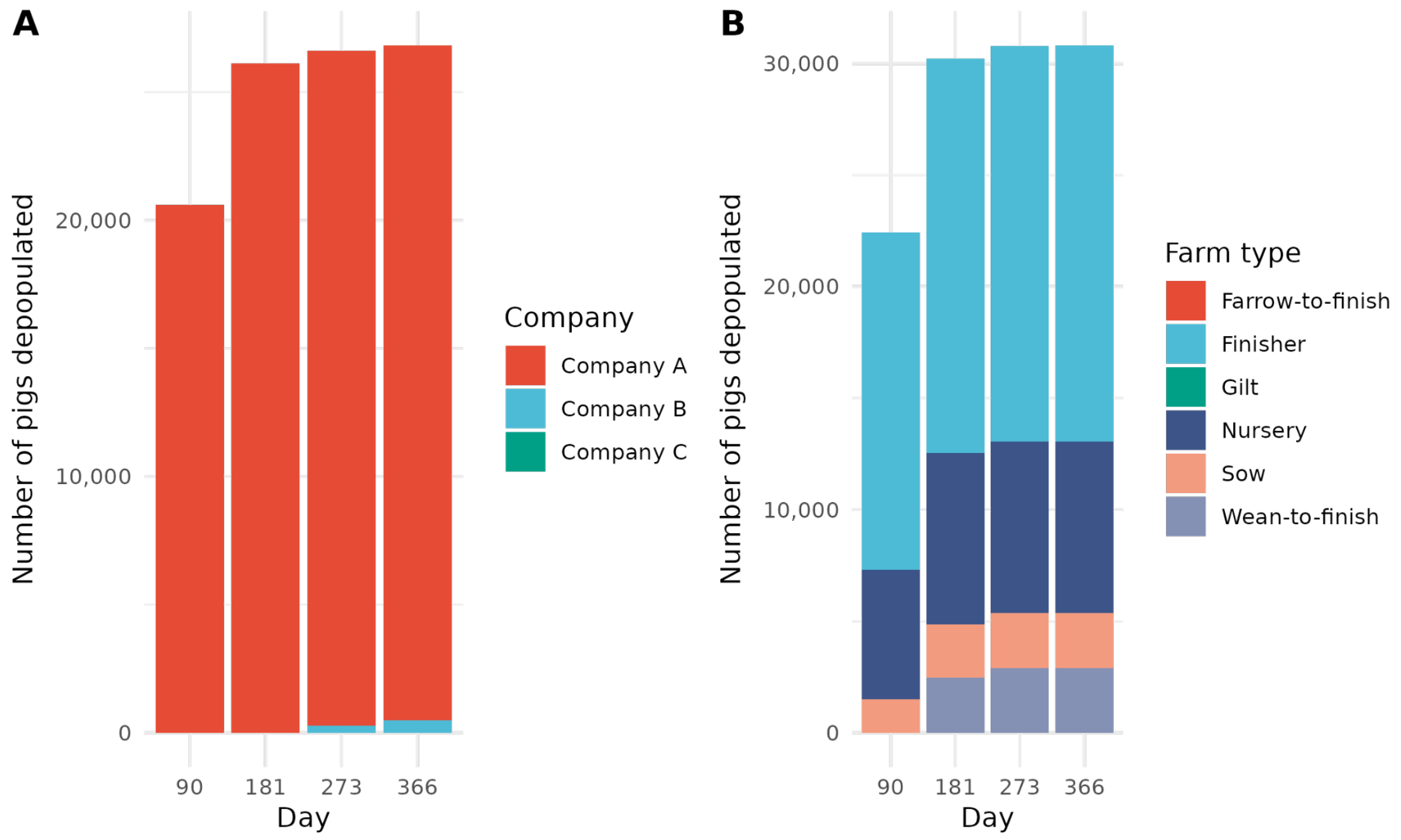

**Figure S7. Median number of depopulated pigs for scenario 2.** A) categorized by companies; B) by farm types

**Scenario 3** involves an extended control zone duration of 40 days, compared to the 30-day duration specified in the NRP (National Response Plan). In this scenario, all other control measures remain consistent with the NRP.

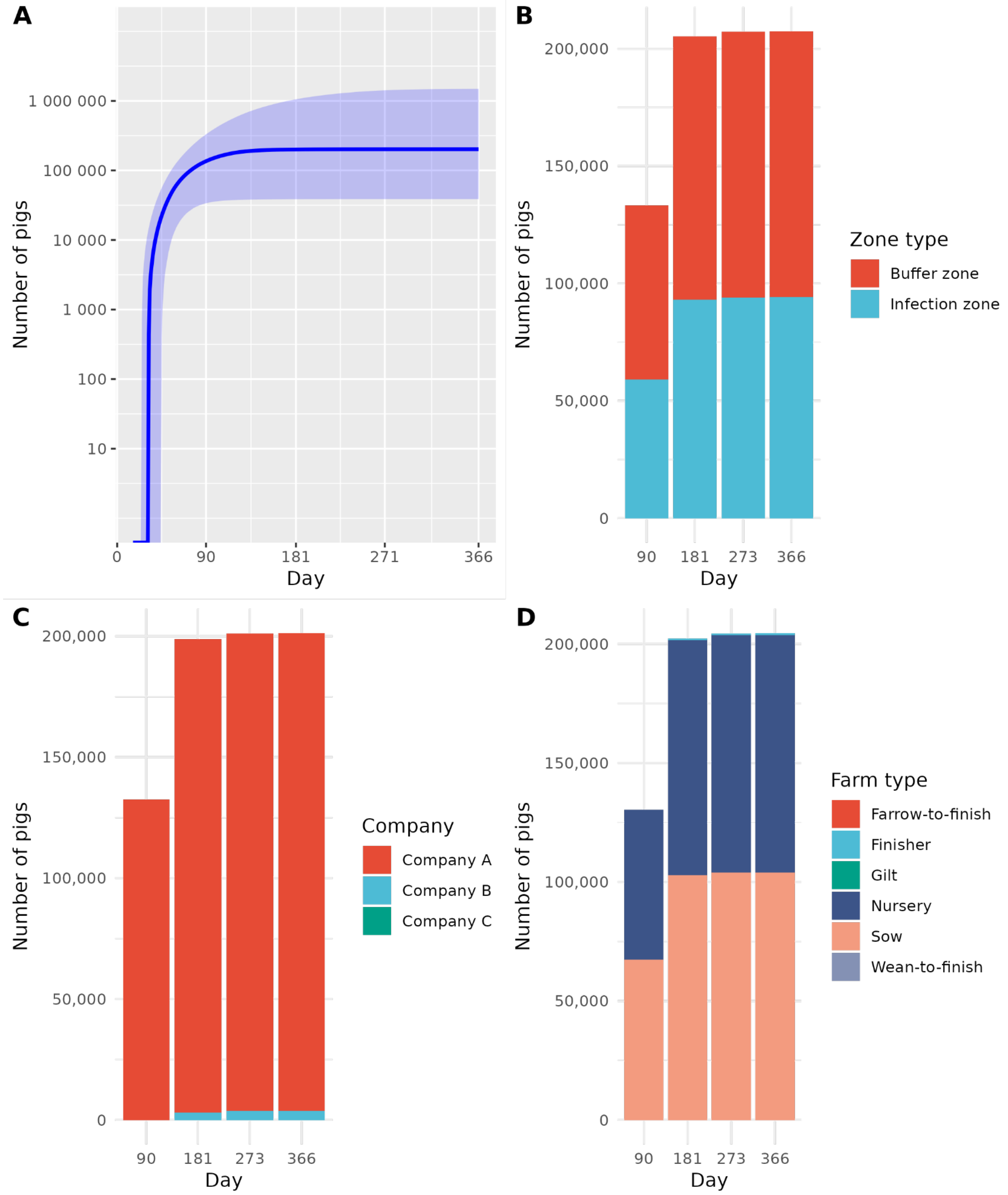

**Figure S8. Cumulative number of unplaced pigs for scenario 3**. A) The daily number of pigs not placed transformed on a logarithmic scale for 366 days, where the solid curve represents the median value, and the shaded area represents the 25% quantile to 75% quantile; B) the median number of pigs not placed categorized by infection and buffer zone type; C) the median number of pigs not placed within control zones categorized by companies; D) the median number of pigs not placed within control zones categorized by origin farm type.

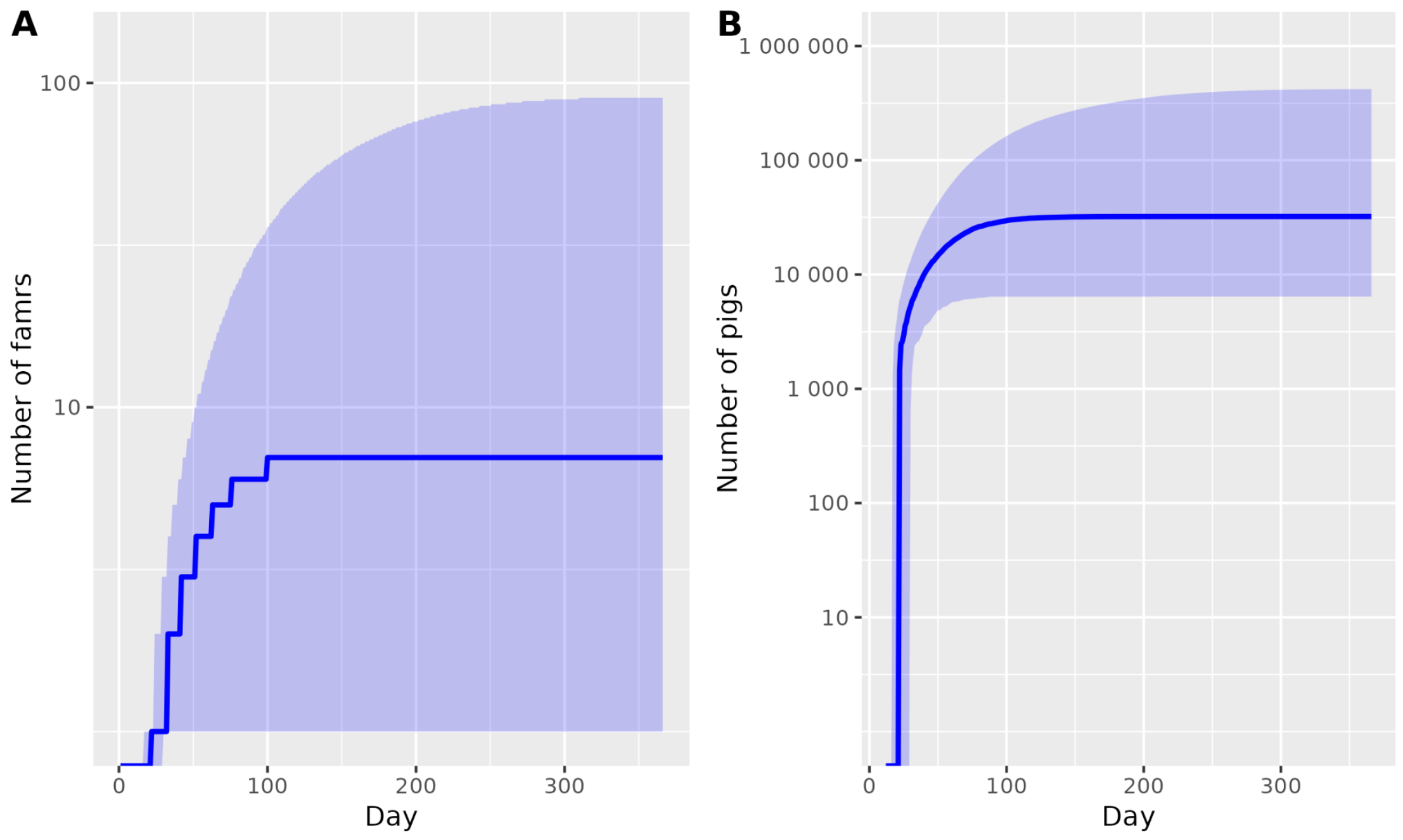

**Figure S9. Cumulative number of infected farms and depopulated pigs for scenario 3.** A) The daily number of infected farms transformed on a logarithmic scale for 366 days; B) The daily number of depopulated pigs transformed on a logarithmic scale for 366 days, where the solid curve represents the median value, and the shaded area represents the 25% quantile to 75% quantile.

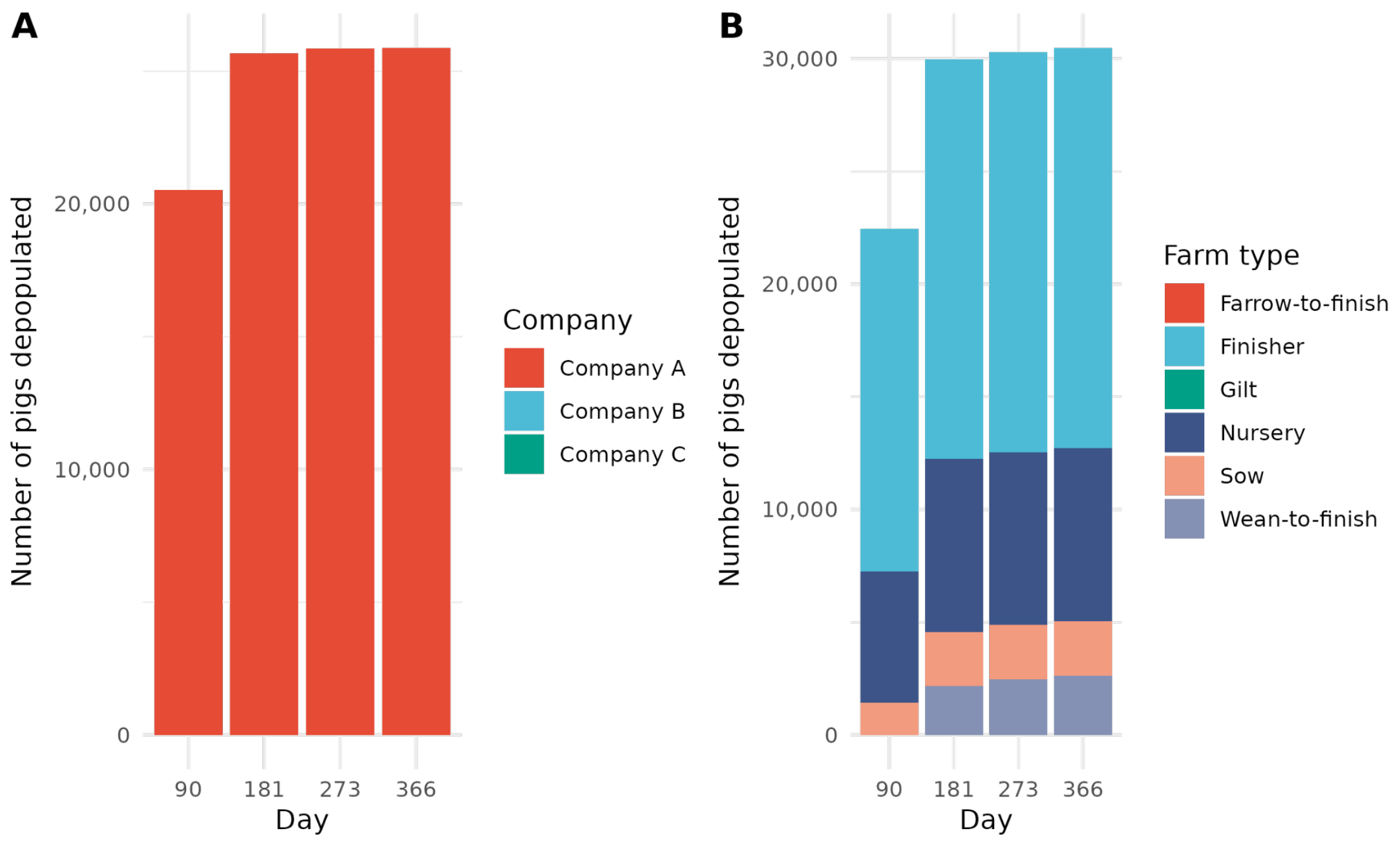

**Figure S10. Median number of depopulated pigs for scenario 3.** A) categorized by companies; B) by farm types

**Scenario 4** involves an expanded infection zone radius of 5 km, compared to the 3 km radius specified in the NRP (National Response Plan). In this scenario, all other control measures remain consistent with the NRP.

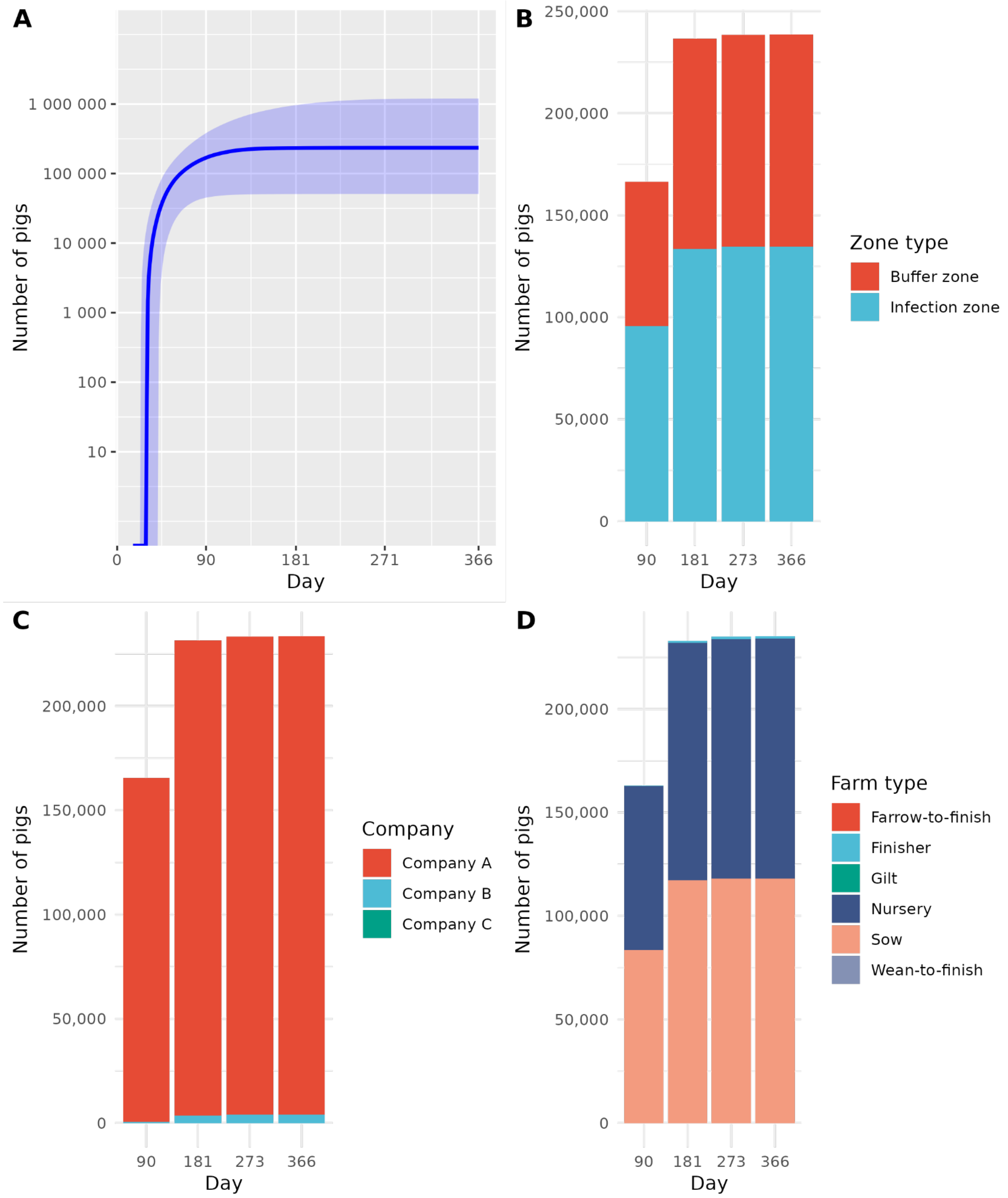

**Figure S11. Cumulative number of unplaced pigs for scenario 4**. A) The daily number of pigs not placed transformed on a logarithmic scale for 366 days, where the solid curve represents the median value, and the shaded area represents the 25% quantile to 75% quantile; B) the median number of pigs not placed categorized by infection and buffer zone type; C) the median number of pigs not placed within control zones categorized by companies; D) the median number of pigs not placed within control zones categorized by origin farm type.

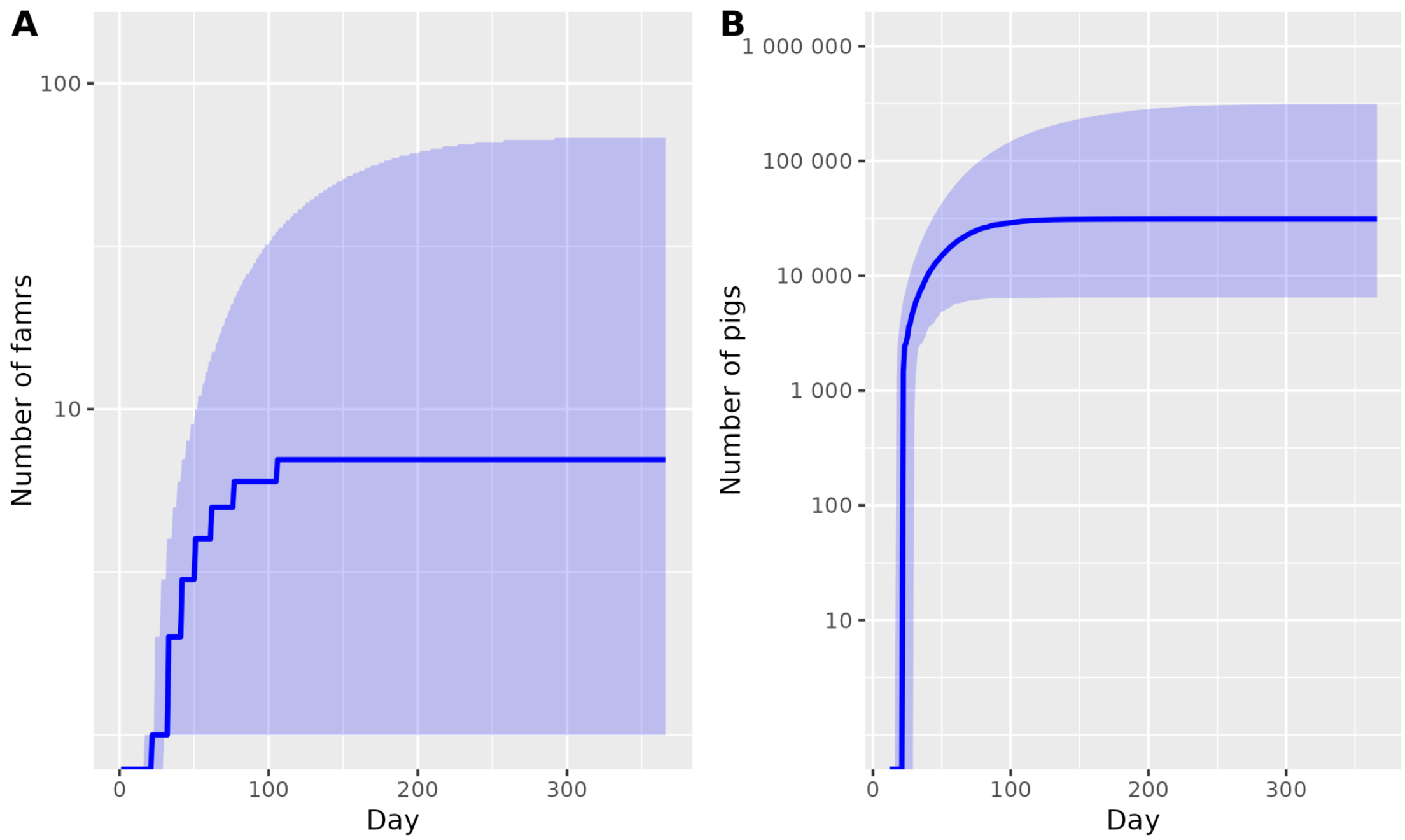

**Figure S12. Cumulative number of infected farms and depopulated pigs for scenario 4.** A) The daily number of infected farms transformed on a logarithmic scale for 366 days; B) The daily number of depopulated pigs transformed on a logarithmic scale for 366 days, where the solid curve represents the median value, and the shaded area represents the 25% quantile to 75% quantile.

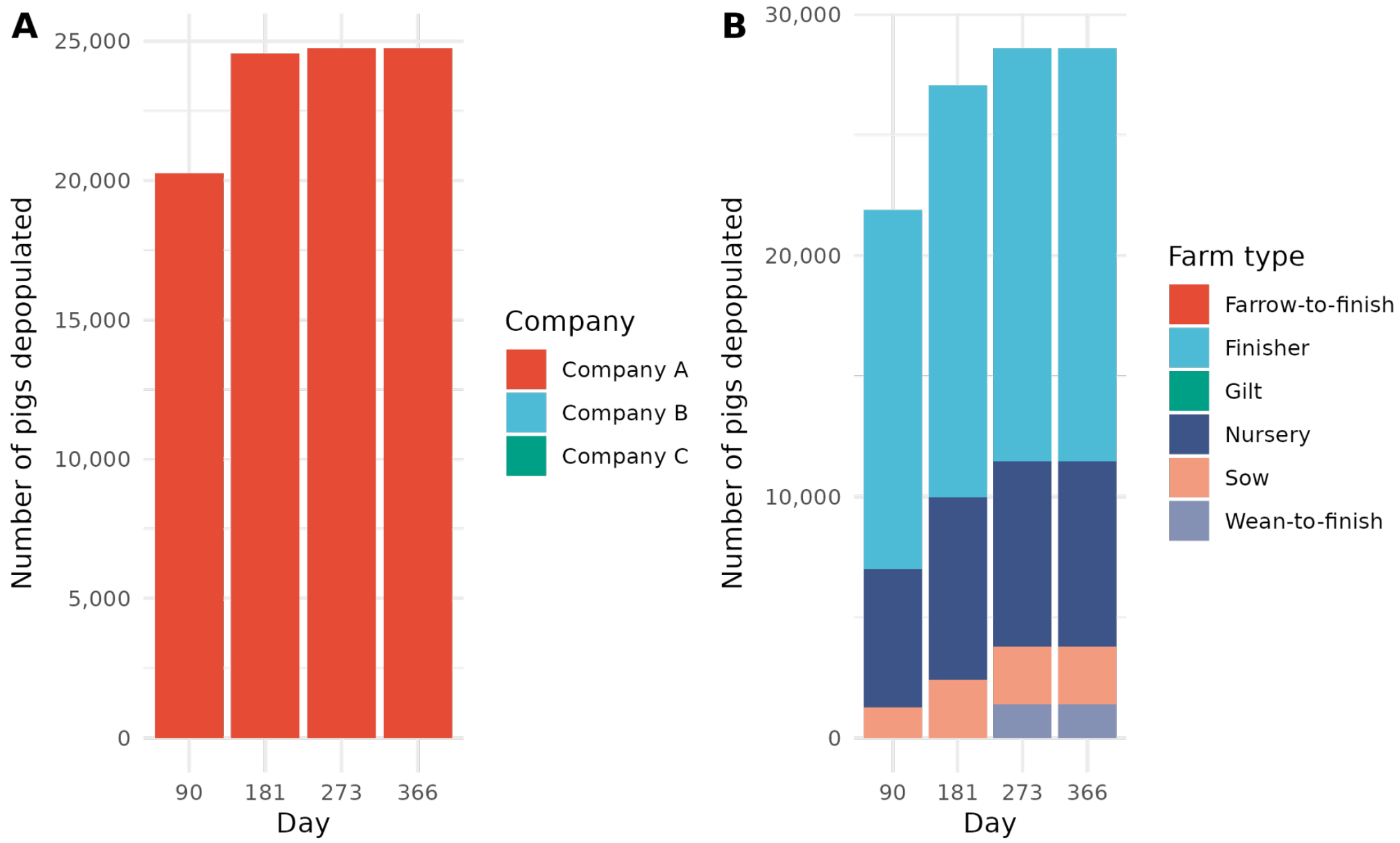

**Figure S13. Median number of depopulated pigs for scenario 4.** A) categorized by companies; B) by farm types

**Scenario 5** involves an expanded infection zone radius of 10 km, compared to the 3 km radius specified in the NRP (National Response Plan). In this scenario, all other control measures remain consistent with the NRP.

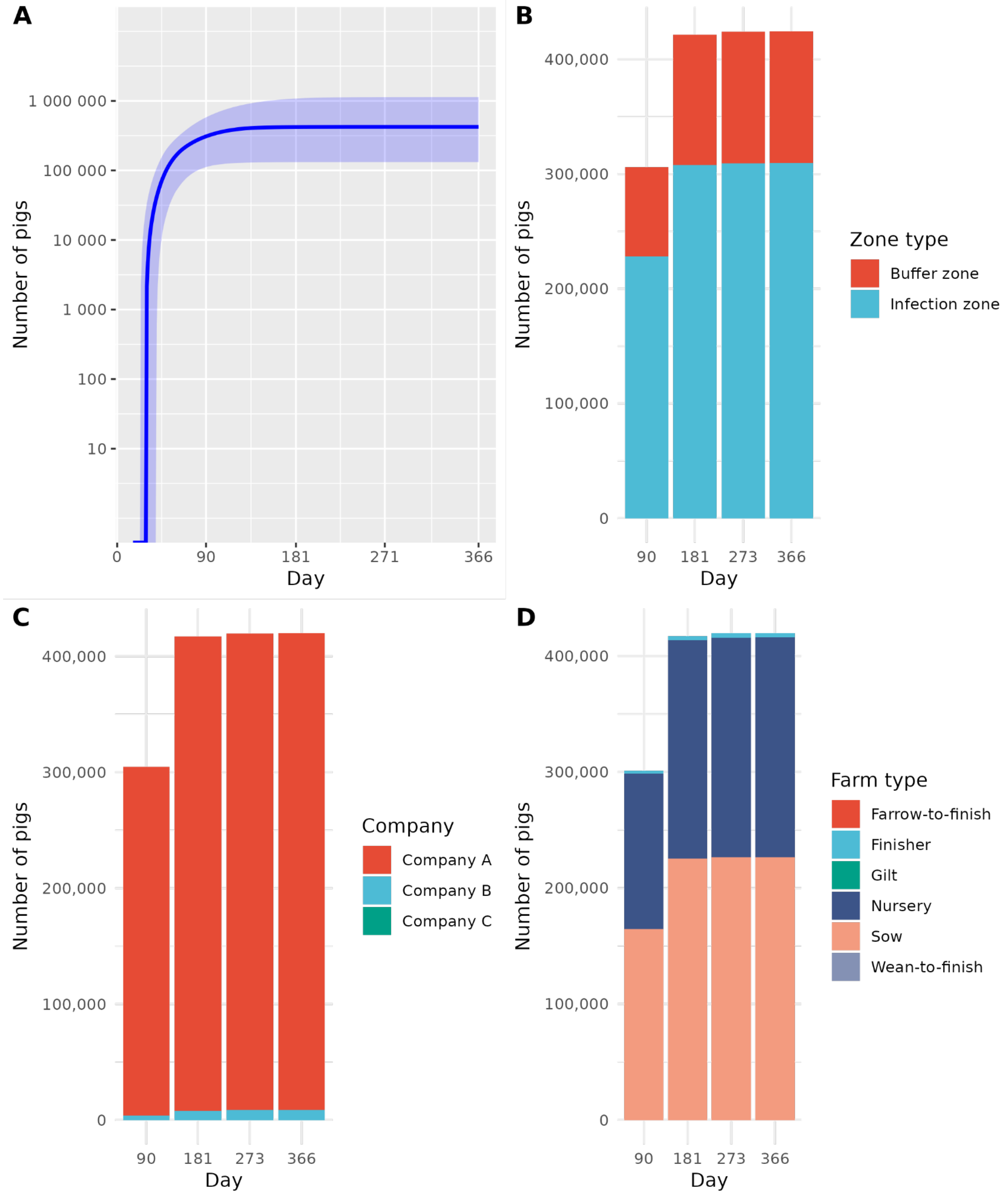

**Figure S14. Cumulative number of unplaced pigs for scenario 5**. A) The daily number of pigs not placed transformed on a logarithmic scale for 366 days, where the solid curve represents the median value, and the shaded area represents the 25% quantile to 75% quantile; B) the median number of pigs not placed categorized by infection and buffer zone type; C) the median number of pigs not placed within control zones categorized by companies; D) the median number of pigs not placed within control zones categorized by origin farm type.

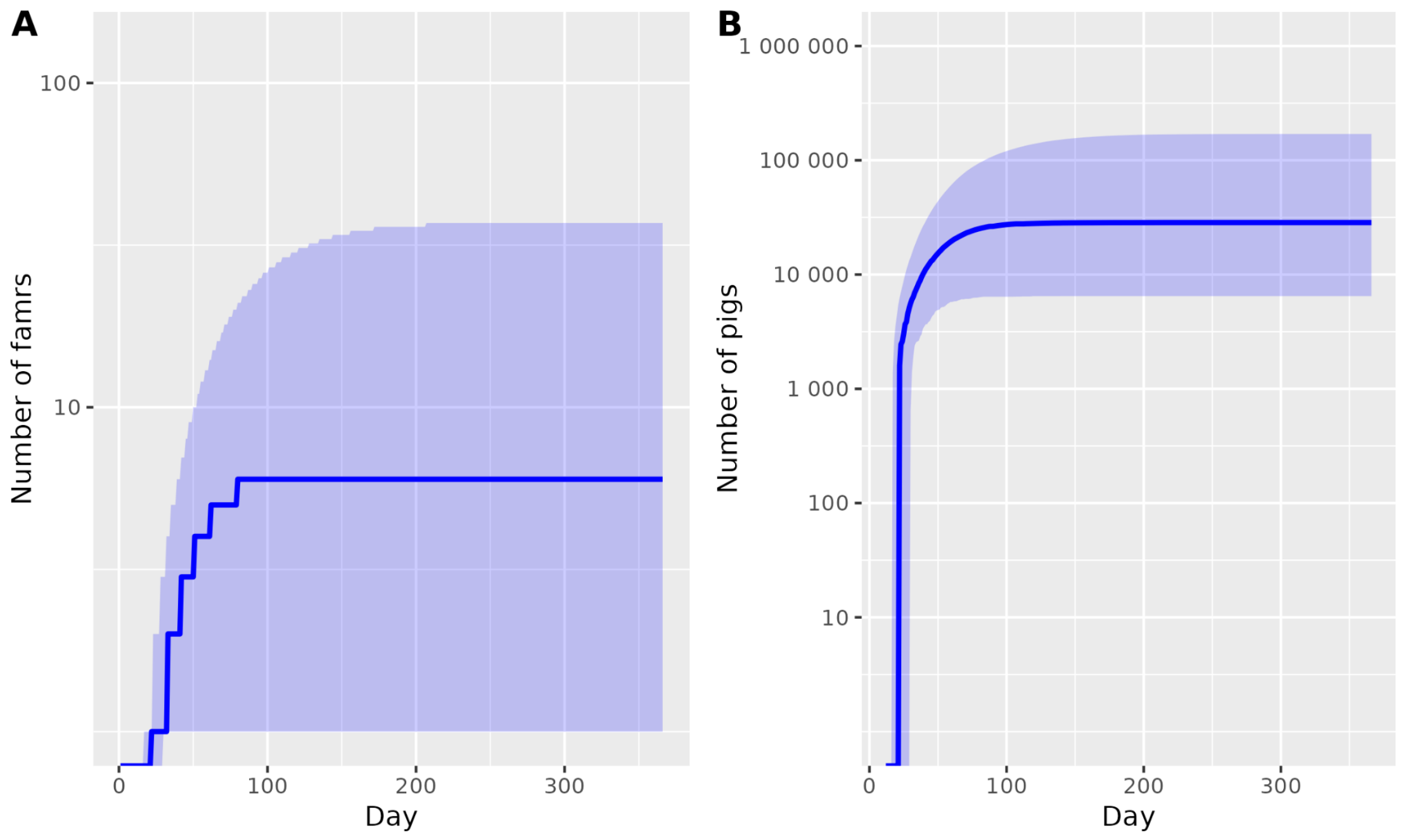

**Figure S15. Cumulative number of infected farms and depopulated pigs for scenario 5.** A) The daily number of infected farms transformed on a logarithmic scale for 366 days; B) The daily number of depopulated pigs transformed on a logarithmic scale for 366 days, where the solid curve represents the median value, and the shaded area represents the 25% quantile to 75% quantile.

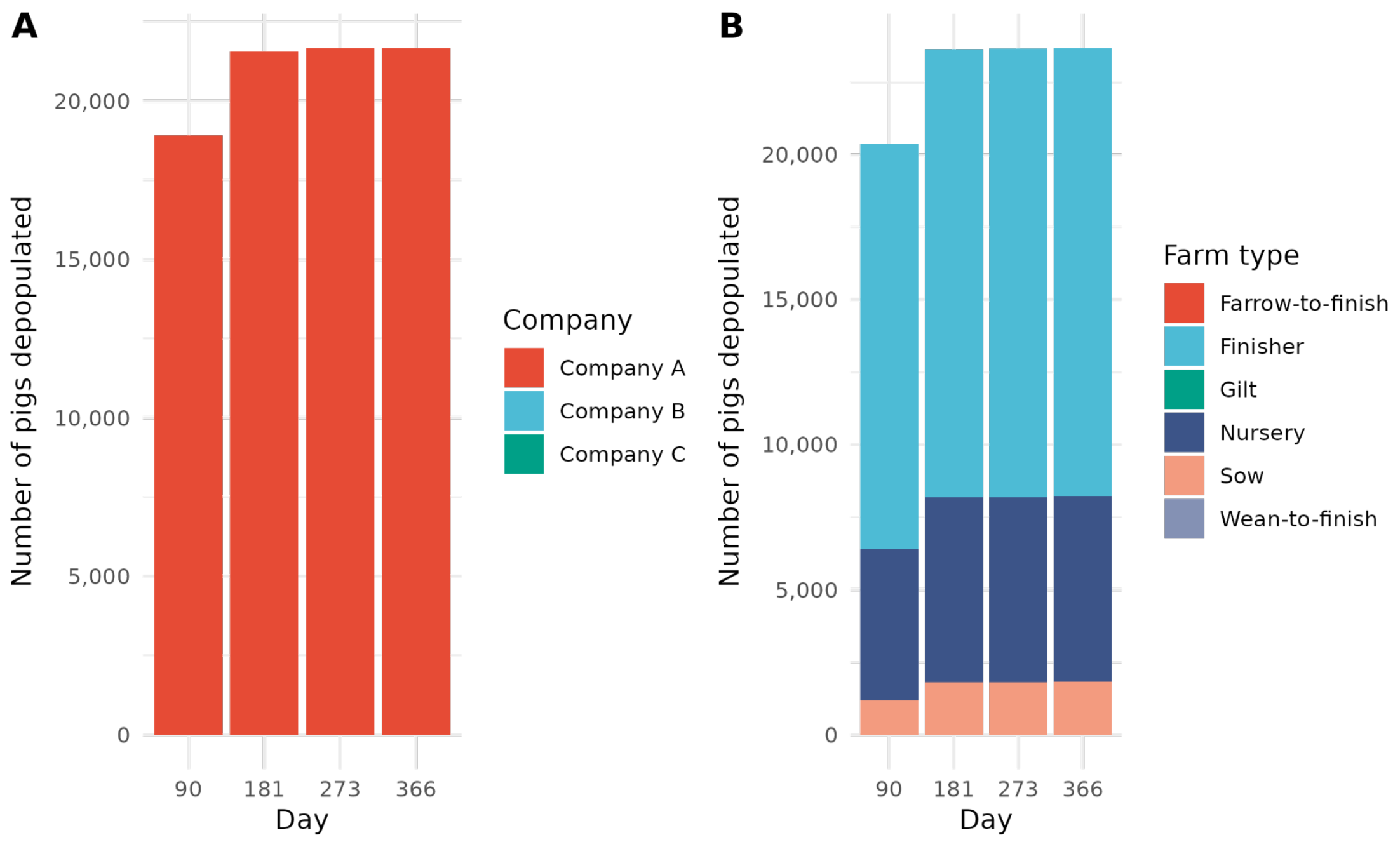

**Figure S16. Median number of depopulated pigs for scenario 5.** A) categorized by companies; B) by farm types

**Scenario 6** involves an expanded infection zone radius of 15 km, compared to the 3 km radius specified in the NRP (National Response Plan). In this scenario, all other control measures remain consistent with the NRP.

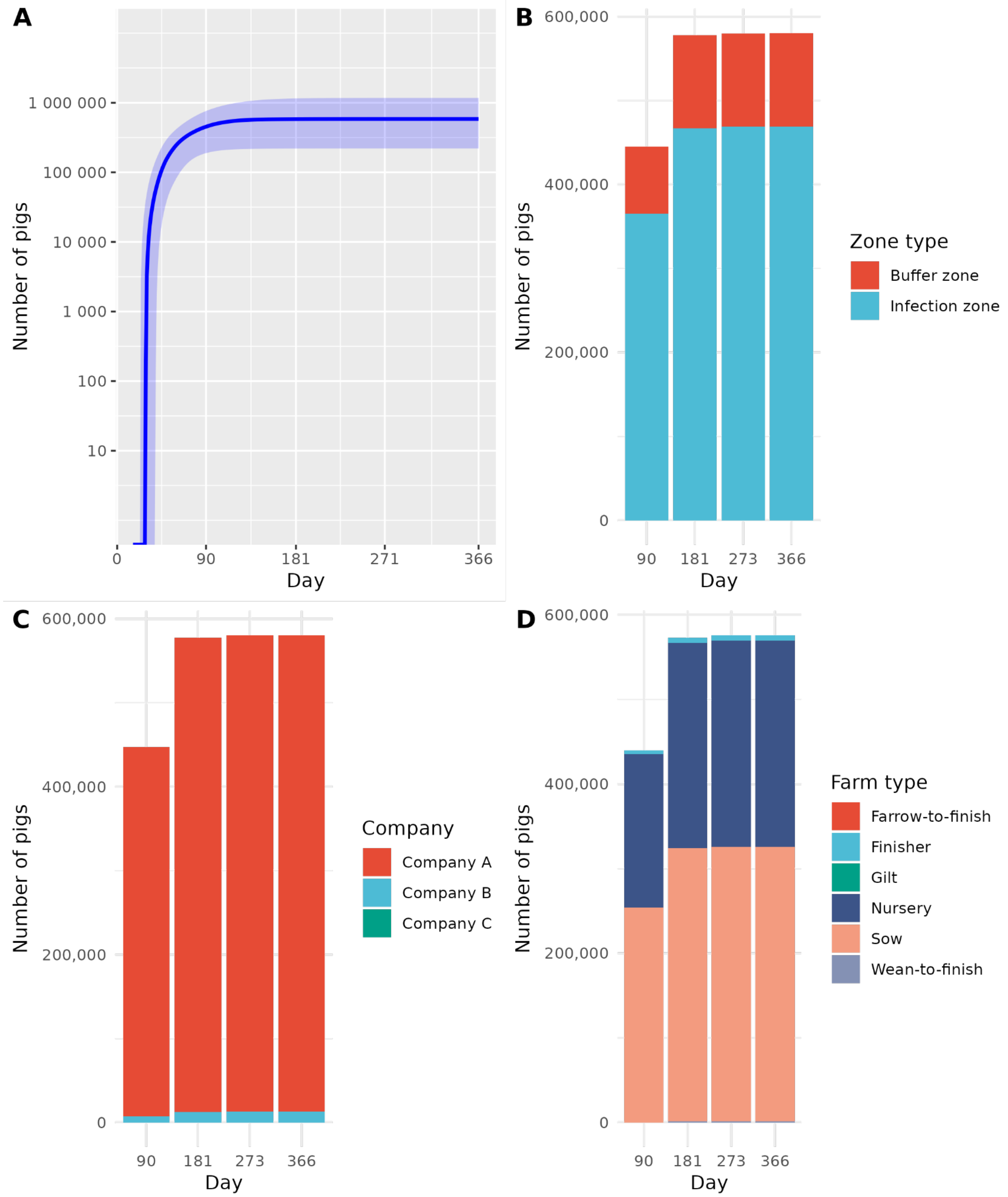

**Figure S17. Cumulative number of unplaced pigs for scenario 6**. A) The daily number of pigs not placed transformed on a logarithmic scale for 366 days, where the solid curve represents the median value, and the shaded area represents the 25% quantile to 75% quantile; B) the median number of pigs not placed categorized by infection and buffer zone type; C) the median number of pigs not placed within control zones categorized by companies; D) the median number of pigs not placed within control zones categorized by origin farm type.

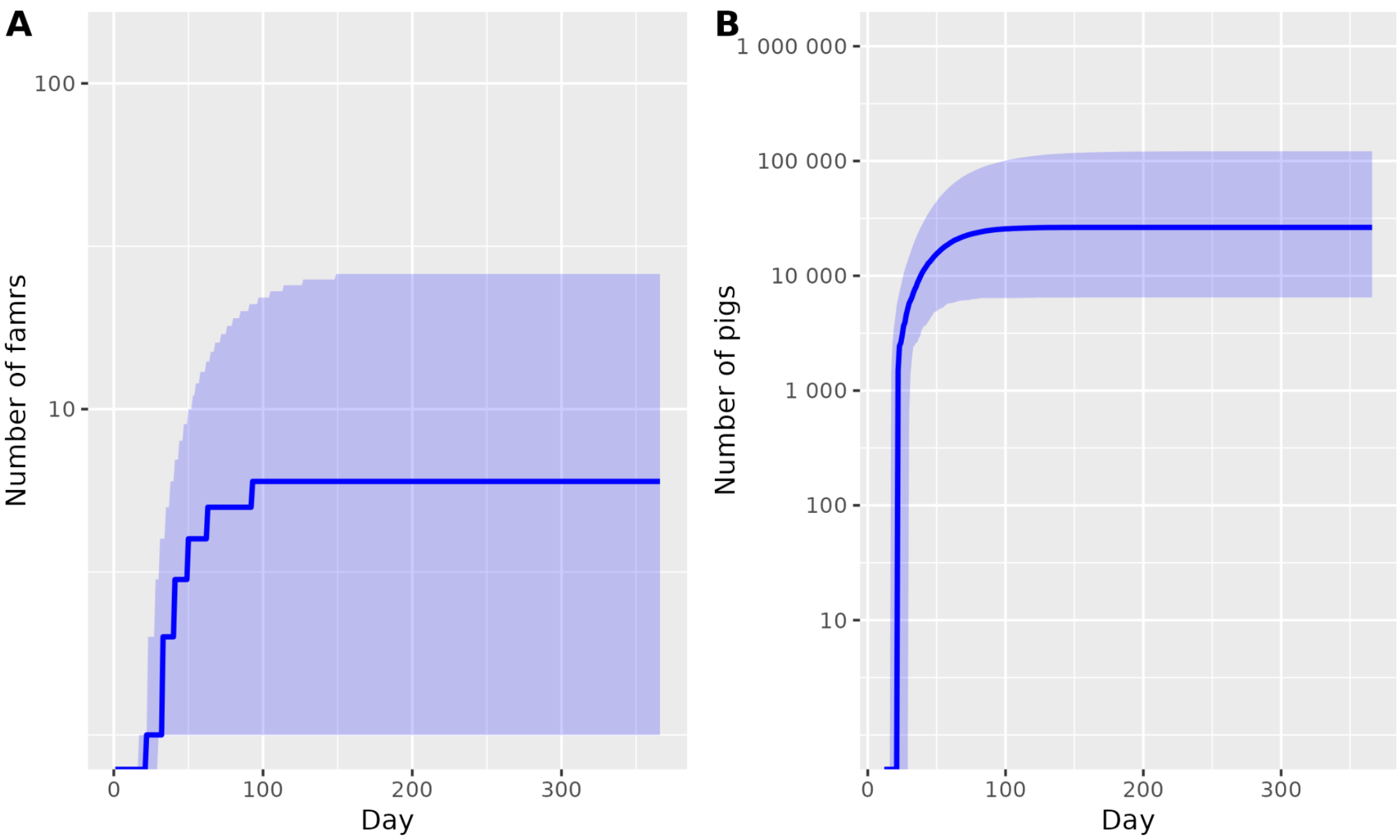

**Figure S18. Cumulative number of infected farms and depopulated pigs for scenario 6.** A) The daily number of infected farms transformed on a logarithmic scale for 366 days; B) The daily number of depopulated pigs transformed on a logarithmic scale for 366 days, where the solid curve represents the median value, and the shaded area represents the 25% quantile to 75% quantile.

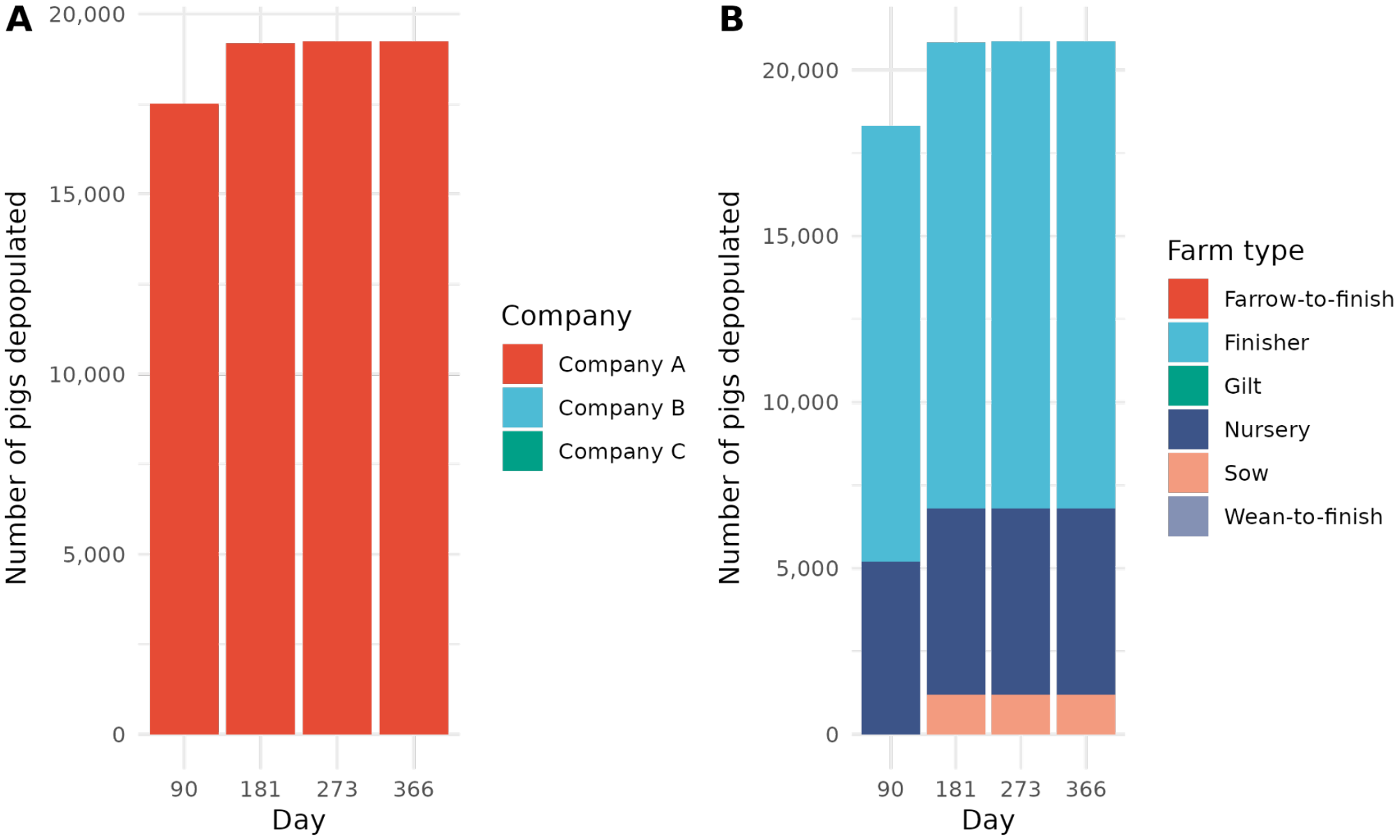

**Figure 19. Median number of depopulated pigs for scenario 6.** A) categorized by companies; B) by farm types

**Calculation of $x$ to minimize the variance of total costs among scenarios.**

The total cost is defined as $L = D + x \cdot U$, where $L, D,$ and $U$ represent total economic loss, the total price of depopulated pigs, and the total price of unplaced pigs, respectively.

Therefore, the median total cost for each scenario is given by:

$L_{NRP} = 6008517 + x \cdot 4199661,$

$L_1 = 6005450 + x \cdot 2783670,$

$L_2 = 6027840 + x \cdot 3488320,$

$L_3 = 6001760 + x \cdot 5539086,$

$L_4 = 5737716 + x \cdot 6417321,$

$L_5 = 5071500 + x \cdot 11{,}580{,}382,$

$L_6 = 4535757 + x \cdot 16058659,$

where those values are median prices from Table 9.

The variance ($\sigma^2$) of the array [$L_{NRP}$, $L_1$, $L_2$, $L_3$, $L_4$, $L_5$, $L_6$] is given by:

$\sigma^2(x) = \frac{\sum_i \quad (L_i - \underline{L})^2}{7}$

where $L_i$ is the total cost of scenario $i$, $\underline{L}$ is the mean cost across scenarios.

By setting the derivative of variance to zero, we can minimize the variance

$\frac{d(\sigma^2(x))}{dx} = 0$ gives $x = 0.12$

**Table S5. Estimated median economic cost $L = D + x \cdot U$ across different scenarios**

| Scenario | NRP | 1 | 2 | 3 | 4 | 5 | 6 |
|---|---|---|---|---|---|---|---|
| $x = 5\%$ | $6,178,310 | $6,101,854 | $6,176,054 | $6,231,816 | $6,027,840 | $5,625,311 | $5,332,335 |
| $x = 12\%$ | $6,457,700 | $6,288,451 | $6,397,825 | $6,635,352 | $6,430,781 | $6,402,238 | $6,399,864 |
| $x = 20\%$ | $6,787,298 | $6,526,727 | $6,667,051 | $7,082,767 | $6,945,725 | $7,363,563 | $7,751,239 |

*In the National Response Plan scenario (**NRP**), control zone duration is 30 days, infection zone radius is 3 km; in **scenario 1**, control zone duration is 20 days, infection zone radius is 3 km; in **scenario 2**, control zone duration is 25 days, infection zone radius is 3 km; in **scenario 3**, control zone duration is 40 days, infection zone radius is 3 km; in **scenario 4**, control zone duration is 30 days, infection zone radius is 5 km; in **scenario 5**, control zone duration is 30 days, infection zone radius is 10 km; in **scenario 6**, control zone duration is 30 days, infection zone radius is 15 km.